\documentclass[letterpaper, 10 pt, conference]{ieeeconf}  
\usepackage{hyperref}
\usepackage{siunitx}
\usepackage{amsmath}
\usepackage{amsfonts}
\usepackage{graphicx}
\makeatletter
\let\MYcaption\@makecaption
\makeatother
\usepackage{caption}
\usepackage{subcaption}

\makeatletter
\let\@makecaption\MYcaption
\makeatother

\usepackage{xcolor}
\usepackage{dblfloatfix}
\usepackage{kantlipsum}
\usepackage{booktabs }
\usepackage{placeins}
\usepackage{float}
\let\labelindent\relax
\usepackage{enumitem}
\usepackage[backend=bibtex,citestyle=numeric-comp,bibstyle=ieee,sorting=none,firstinits=true,doi=false,url=false,isbn=false,date=year]{biblatex}
 \usepackage{physics}
 \usepackage{mathtools, cuted}
 \usepackage{flushend}

\newtheorem{thm}{Theorem}     

\IEEEoverridecommandlockouts                              
\title{\LARGE \bf Beyond the Waterbed Effect: Development of Fractional Order \\CRONE Control with Non-Linear Reset
}

\author{Linda Chen,  Niranjan Saikumar, Simone Baldi and S. Hassan HosseinNia
\thanks{Linda Chen and Simone Baldi are with the Delft Center for Systems and Control, Delft University of Technology, Delft, The Netherlands
        {\tt\small lindachen93@gmail.com, S.Baldi@tudelft.nl}}%
\thanks{Niranjan Saikumar and S. Hassan HosseinNia are with the Department of Precision and Microsystems Engineering, Delft University of Technology, Delft, The Netherlands
        {\tt\small N.Saikumar@tudelft.nl, S.H.HosseinNiaKani@tudelft.nl}}%
}

\bibliography{reference.bib}

\begin{document}

\maketitle
\thispagestyle{empty}
\pagestyle{empty}

\begin{abstract}
In this paper a novel reset control synthesis method is proposed: CRONE reset control, combining a robust fractional CRONE controller with non-linear reset control to overcome waterbed effect. In CRONE control, robustness is achieved by creation of constant phase behaviour around bandwidth with the use of fractional operators, also allowing more freedom in shaping the open-loop frequency response. However, 
being a linear controller it suffers from the inevitable trade-off between robustness and performance as a result of the waterbed effect.
Here reset control is introduced in the CRONE design to overcome the fundamental limitations. 
In the new controller design, reset phase advantage is approximated using describing function analysis and used to achieve better open-loop shape.
Sufficient quadratic stability conditions are shown for the designed CRONE reset controllers and the control design is validated on a Lorentz-actuated nanometre precision stage. It is shown that  for similar phase margin, better performance in terms of reference-tracking
and noise attenuation can be achieved.
\end{abstract}

\section{Introduction}
%
%
%
%

In high-tech industry not only (sub)nanometre precision positioning control, but also high bandwidth is essential: this is the case for wafer scanners used for production of integrated circuits, atomic force microscopes for scanning of dynamic biological samples and printing of three-dimensional nano-structures. 
It is in demanding cases such as these that requirements for robustness begin to conflict with requirements for reference-tracking, disturbance rejection and noise attenuation, as a result of a fundamental trade-off between robustness and performance. This trade-off is well-studied in the waterbed effect \cite{skogestad2007multivariable} and considered in precision positioning system design \cite{tan2007precision},\cite{schmidt2014design}. 

Loop shaping is the industry-standard for control design which allows for performance evaluation in frequency domain. High open-loop gain at low frequency is required for better tracking and effective rejection of disturbances (like external vibrations) and low gain is required at high frequency for precision with better noise attenuation. On the other hand, better robustness of the system is achieved for more phase around bandwidth. 
Bode's gain-phase relation imposes a direct constraint between the gain and phase behaviour. Thus if the phase is increased in the frequency range around bandwidth as to increase robustness, low-frequency open-loop gain has to decrease and high-frequency open-loop gain has to increase accordingly, degrading performance of the system. PID controllers, which have been an industry-standard for many years, do not satisfy the ever increasing demands on robustness and performance. By using a fractional order controller such as CRONE \cite{sabatier2015fractional}, additional tunability allows for improved controller design. However, fundamental limitations of linear control remain. This motivates the use of non-linear controllers as a way to break the aforementioned trade-offs.  

A promising non-linear control technique in this sense is reset control. J.C. Clegg  introduced the reset integrator in 1958 and showed using describing function analysis \cite{vidyasagar2002nonlinear} that a 52\si{\degree} phase lead is achieved when compared to a linear integrator for the same \si{-20\deci\bel/}decade slope. 
The improvement in performance of linear systems using reset control has been shown in for instance \cite{prieur2011} and \cite{witvoet2007} amongst others. Today, numerous works on reset control theory exist, with practical applications ranging from: hard-drive disk control \cite{Li2011}, \cite{Li2009}, servomotor control \cite{hassan2013}, positioning stages \cite{hazeleger2016}, \cite{zheng2007} and process control \cite{davo2013}, \cite{perez2011}, \cite{moreno2013}. Reset control design and stability of reset systems remain actively researched topics today. 
A few results include the PI+CI compensator (PI controller with a Clegg integrator) for which a control design framework has been developed in \cite{banos2011reset} and time-regularized reset controllers for which asymptotic stability has been guaranteed in \cite{banos2011}. In a recent work, sufficient stability conditions based on measured frequency responses are given \cite{van2017frequency}, which aims to elevate the need for solving linear matrix inequalities (LMI) and thus making reset controllers more accessible to control engineers in industry. 

Despite the advances mentioned above, a full and accessible framework for reset control synthesis is to the best of the authors' knowledge, still an open problem. In order to improve tunability in reset control, generalization of reset control to fractional order has already been addressed in \cite{hassan2014} and \cite{niranjan2017}. The aim of this research is to incorporate fractional reset theory developed in mentioned works into CRONE design methodology. The novel CRONE reset controller, could simplify reset control design by enabling the use of common loop shaping methods. Motivation of this research direction arises from the possibility to break the fundamental robustness performance trade-off on the one hand and simplifying robust reset control design on the other.


The rest of the paper is organized as follows: background information about CRONE control and reset control is given in section \ref{sec:background}. In section \ref{sec:cronereset} the proposed CRONE reset controller is discussed, followed by a stability analysis of the CRONE reset controller provided in section \ref{sec:stability}. Validation of the CRONE reset controller on an experimental setup is given in section \ref{sec:validation} followed by a discussion of results in section \ref{sec:results} and conclusions in section \ref{sec:conclusion}.

\section{Background}\label{sec:background}
\subsection{CRONE control}
\begin{figure}
\centering
\begin{subfigure}{0.45\linewidth}
\flushleft
\includegraphics[width=\linewidth]{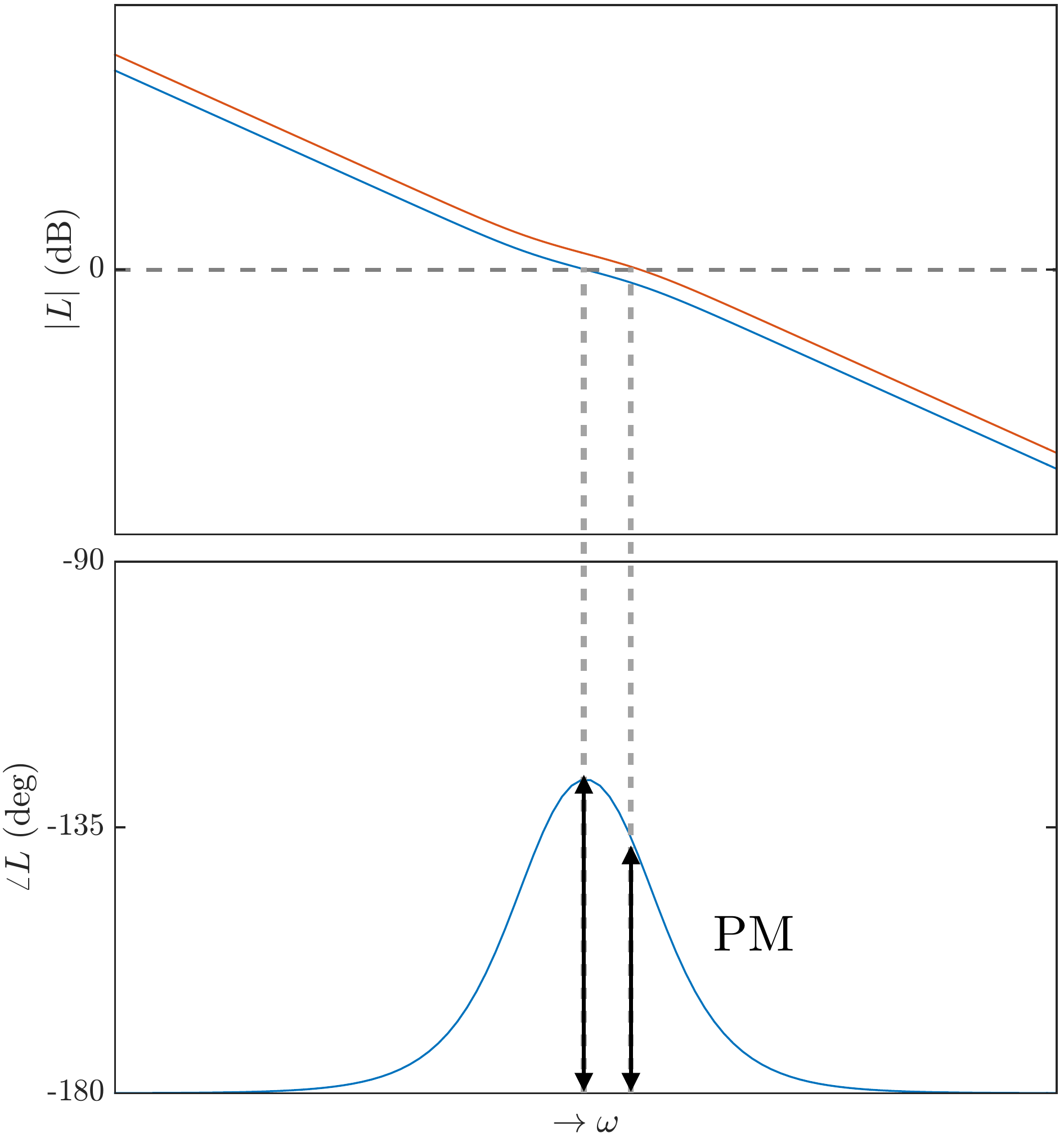}
\subcaption{}
\label{img:bodecrone1}
\end{subfigure}
\begin{subfigure}{0.45\linewidth}
\centering
\includegraphics[width=\linewidth]{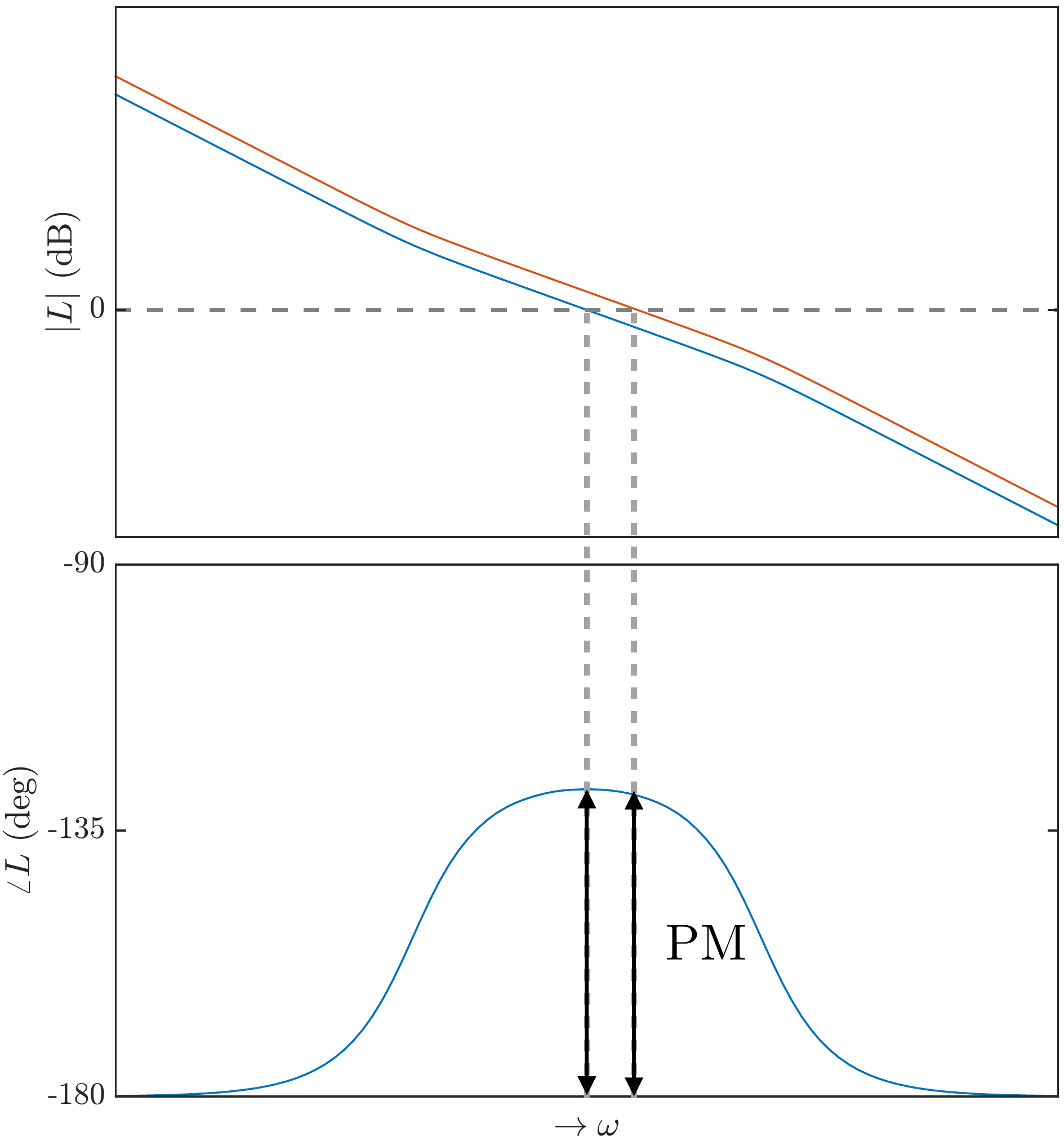}
\subcaption{}
\label{img:bodecrone2}
\end{subfigure}
\caption{(\subref{img:bodecrone1}) Typical open-loop response. Under gain variations the phase margin (PM) fluctuates. (\subref{img:bodecrone2}) After achieving constant phase in the frequency range around bandwidth the phase margin is constant.}
\label{img:bodecrone}
\end{figure}

In CRONE control robustness against gain deviations in the system is obtained by creating constant phase behaviour around bandwidth in open-loop. This can be seen in Fig. \ref{img:bodecrone}. Three generations of CRONE exist today as formalized by \cite{sabatier2015fractional}. Our focus is on first generation CRONE control (CRONE-1), which can be used for plants with asymptotic phase around bandwidth.

A first generation CRONE controller has a similar transfer function to an integer order series PID controller:
\begin{equation}\label{eq:Cf}
C_F(s)=C_0(1+\frac{\omega_I}{s})^{n_I}\bigg(\frac{1+\frac{s}{\omega_b}}{1+\frac{s}{\omega_h}}\bigg)^\nu\frac{1}{(1+\frac{s}{\omega_F})^{n_F}}
\end{equation}

with $\omega_I$ and $\omega_F$ being the integrator- and low pass filter corner frequencies, $\omega_b$ and $\omega_h$ the corner frequencies of the band-limited derivative action, $\nu \in \mathbb{R} \cap [0,1]$ the fractional order and $n_I,n_F \in \mathbb{N} $ being the order of the integrator and low pass filter order respectively. The difference between a series integer order PID controller and a first generation CRONE controller is that the order $\nu$ is fractional instead of integer, making first generation CRONE a fractional PID controller. The fractional part of the transfer function can be approximated using Oustaloup approximation within the frequency range. The flat phase behaviour illustrated in Fig. \ref{img:bodecrone2} is created by choosing a wider frequency range in which the derivative action is active (compared to PID control) and by decreasing the order $\nu$ to a fractional value. This ensures that the gain at high and low frequencies are not affected by the constant phase achieved.


The fractional order $\nu$ can be calculated from:

\begin{multline}\label{eq:alpha}
\nu=\frac{-\pi+M_\Phi-\arg G(j\omega_{cg})+n_F\arctan\frac{\omega_{cg}}{\omega_F}}{\arctan\frac{\omega_{cg}}{\omega_b}-\arctan\frac{\omega_{cg}}{\omega_h}}\\+\frac{n_I(\frac{\pi}{2}-\arctan \frac{\omega_{cg}}{\omega_I})}{\arctan\frac{\omega_{cg}}{\omega_b}-\arctan\frac{\omega_{cg}}{\omega_h}}
\end{multline}
where $M_\Phi$ is the required nominal phase margin, $\omega_{cg}$ the bandwidth, $G(j\omega)$ is the plant frequency response.

The gain $C_0$ is chosen such that the loop gain at frequency $\omega_{cg}$ is equal to 1. 
%
%

\subsection{Reset control}

\begin{figure}[!htb]
\centering
\includegraphics[width=.92\linewidth]{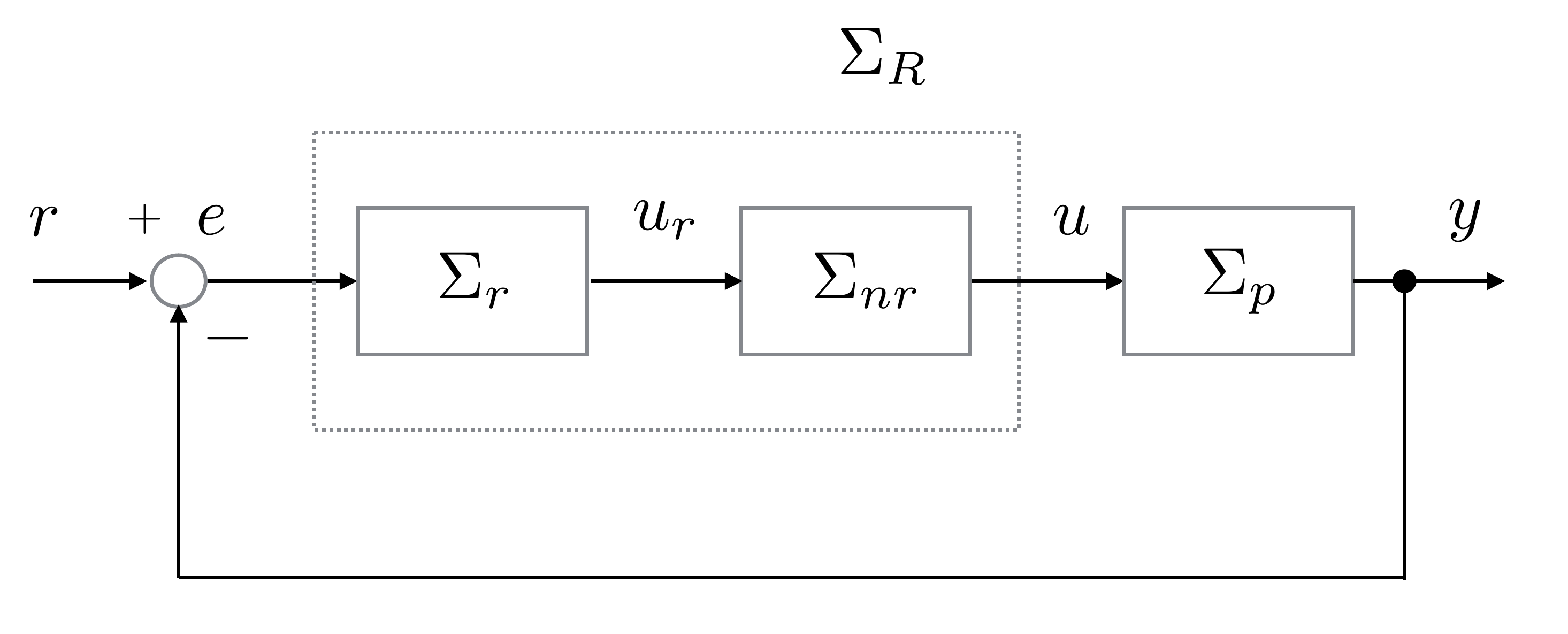}
\caption{Control diagram of feedback control with a controller $\Sigma_R$, which contains a linear system $\Sigma_{nr}$ and a  non-linear reset system $\Sigma_r$.}
\label{fig:rcontrol}
\end{figure}

A general reset controller can be described by the following impulsive differential equations, using the formalism in \cite{banos2011reset}:
\begin{equation}\label{eq:IDESore}
\Sigma_r:=\begin{cases}
\dot{x}_r(t)=A_rx_r(t)+B_re(t)&\text{if } e(t)\neq 0,\\
x_r(t^+)=A_\rho x_r(t)&\text{if } e(t)=0,\\
u(t)=C_rx_r(t)+D_re(t)& \\
\end{cases}
\end{equation}
where matrices $A_r, B_r, C_r, D_r$ are the base linear state-space matrices of the reset controller, $e(t)$ is the error between output and reference, $u(t)$ is the control input signal, $x_r(t)$ is the state and $A_\rho$ is the reset matrix.  A typical control structure using reset control can be seen in Fig. \ref{fig:rcontrol}. The overall controller $\Sigma_R$ consists of a reset controller $\Sigma_r$ and a linear controller $\Sigma_{nr}$. Typical reset elements include the Clegg integrator and the first order reset element (FORE) as in \cite{chen2001}. In the design of a CRONE reset controller, key is to quantify the phase advantage that reset provides: to this purpose, the describing function computation of reset control systems is explained following the formalism of \cite{guo2009frequency}. 

\subsubsection*{Describing function}
The general describing function of a reset system as defined in \cite{guo2009frequency} is given by:
\begin{equation}\label{eq:DF}
G_\mathrm{DF}(j\omega)=C_r(j\omega I-A_r)^{-1}B_r(I+j\Theta_D(\omega))+D_r
\end{equation}
with $\Theta_D(\omega)$:
\begin{equation}\label{eq:thetad}
\Theta_D(\omega)=-\frac{2\omega^2}{\pi}\Delta(\omega)[\Gamma_D(\omega)-\Lambda^{-1}(\omega)]
\end{equation}
in which the following set of equations are present:
\begin{eqnarray*}
\begin{cases}
\Lambda(\omega)=\omega^2I+A^2_r\\
\Delta(\omega)=I+e^{\frac{\pi}{\omega}A_r}\\
\Delta_D(\omega)=I+A_\rho e^{\frac{\pi}{\omega}A_r}
\end{cases}
\end{eqnarray*}

%
%
%

\subsubsection*{Creating phase advantage with reset}
\begin{figure}[!htb]
\centering
\begin{subfigure}{0.47\linewidth}
\includegraphics[width=\linewidth]{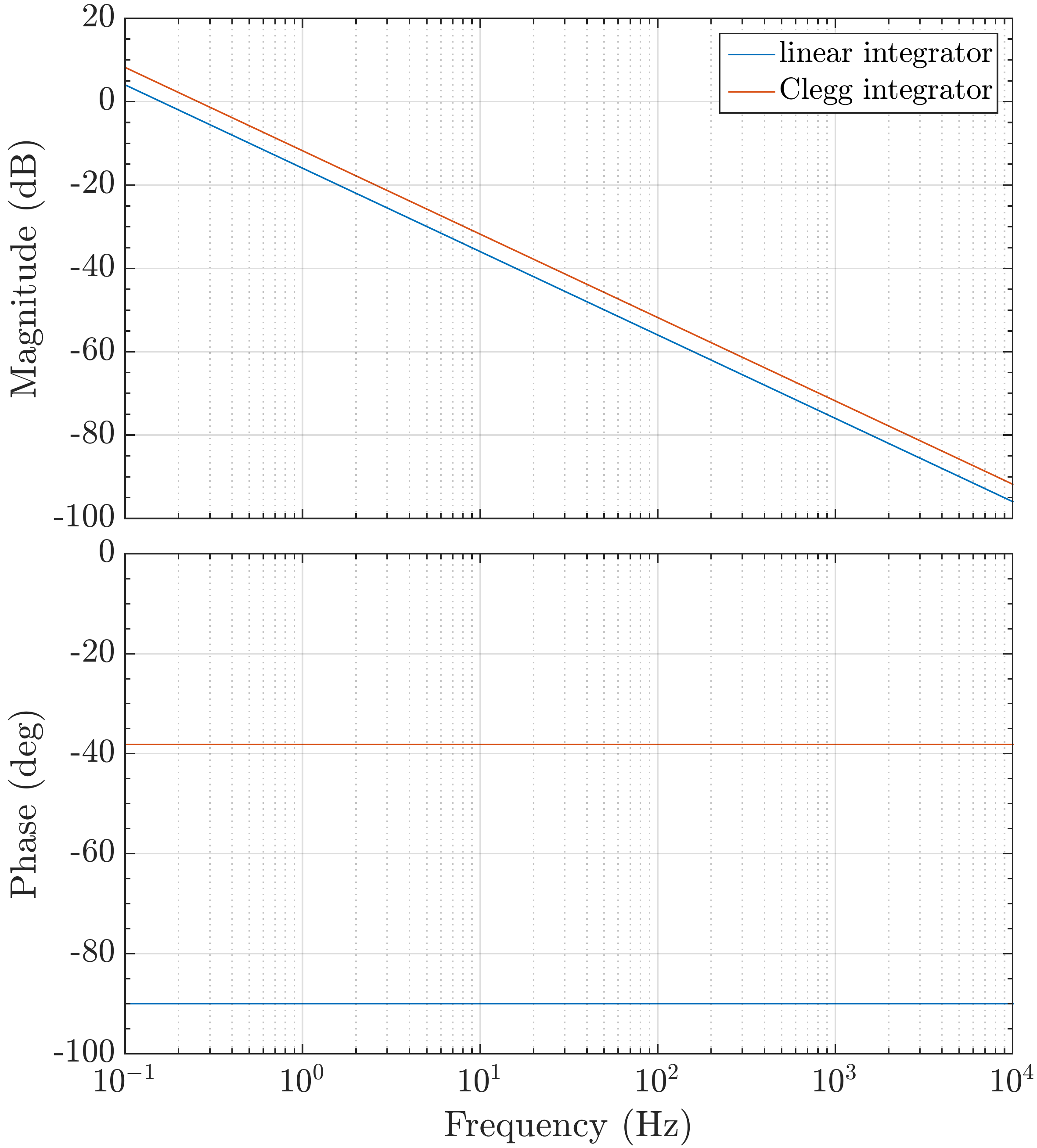}
\caption{}
\label{fig:resetphaseci}
\end{subfigure}
\begin{subfigure}{0.47\linewidth}
\includegraphics[width=\linewidth]{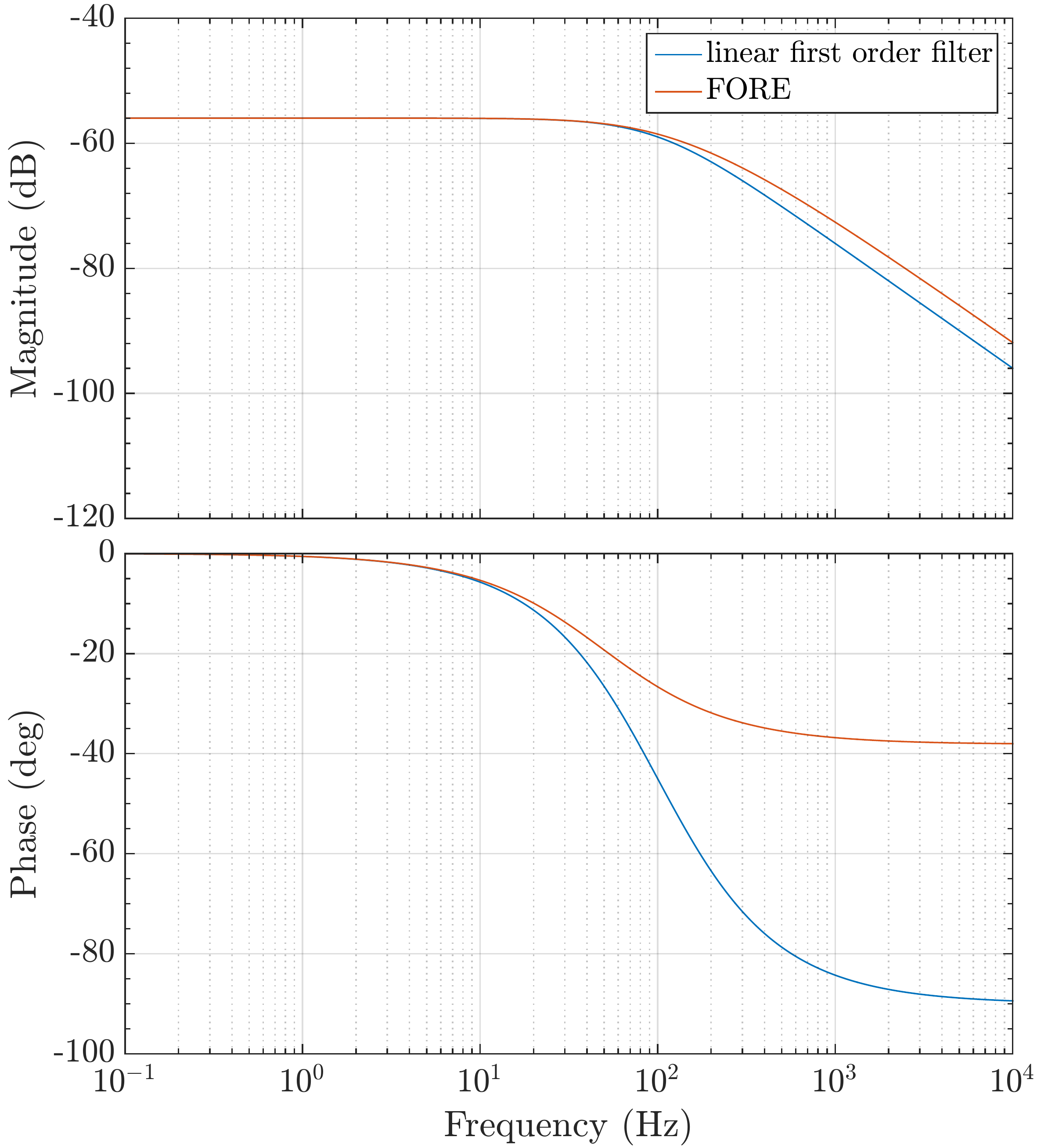}
\caption{}
\label{fig:resetphasefore}
\end{subfigure}
\caption{Describing function of a (\subref{fig:resetphaseci}) Clegg integrator and (\subref{fig:resetphasefore}) FORE compensator showing less phase lag compared to their linear base transfer functions.}
\end{figure}

Using describing function analysis phase and gain behaviour of reset filters can be approximated. Gain behaviour of reset filters very closely resembles that of its linear base filter, whereas phase behaviour can differ significantly. In Fig. \ref{fig:resetphaseci}, the describing function of the Clegg integrator is compared to the frequency response of a linear integrator. In Fig. \ref{fig:resetphasefore} the describing function of a FORE compensator is compared to the frequency response of a first order filter with the same corner frequency of \si{100\hertz}. For the reset integrator phase lead is created for all frequencies, whereas for the FORE reduced phase lag is seen after the corner frequency. 

\section{CRONE reset control}\label{sec:cronereset}
At the base of CRONE reset control is the robust CRONE controller with flat phase behaviour around the bandwidth frequency. With reset action additional phase can be created around bandwidth. When combining CRONE with reset, these two situations would allow for decrease of the open-loop slope around bandwidth by retuning fractional order $\nu$. This has beneficial effects on system performance, as the low-frequency loop gain increases and the high-frequency loop gain decreases as a result. The parameters chosen in the control structure determine the amount of additional phase created by reset. This phase lead is calculated using the describing function and used in the computation of the new fractional order $\nu$. In the subsections below firstly the control structure is explained and then an adapted formula for the fractional order is given.

\begin{figure}[!htb]
\centering
\begin{subfigure}{\linewidth}
\centering
\includegraphics[width=\linewidth]{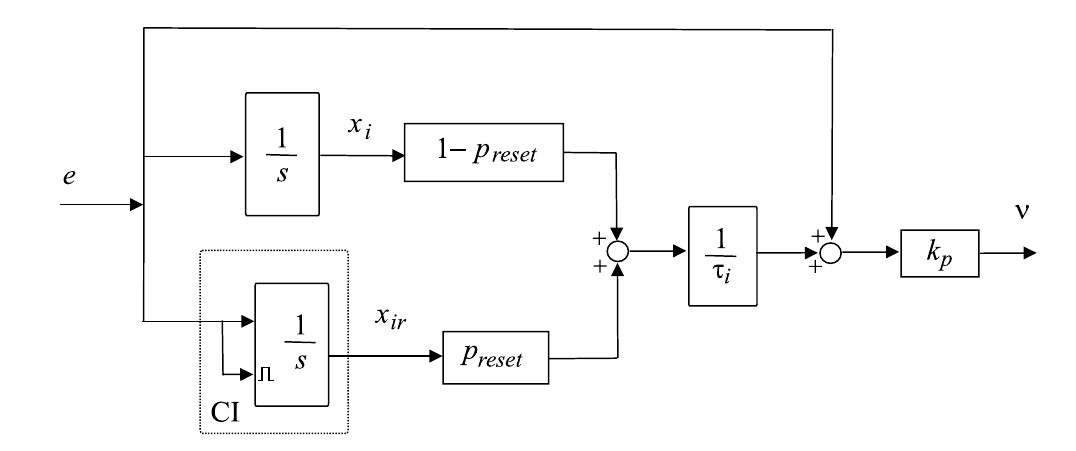}
\caption{}
\label{fig:PI+CI}
\end{subfigure}\\
\begin{subfigure}{\linewidth}
\centering
\includegraphics[width=.85\linewidth]{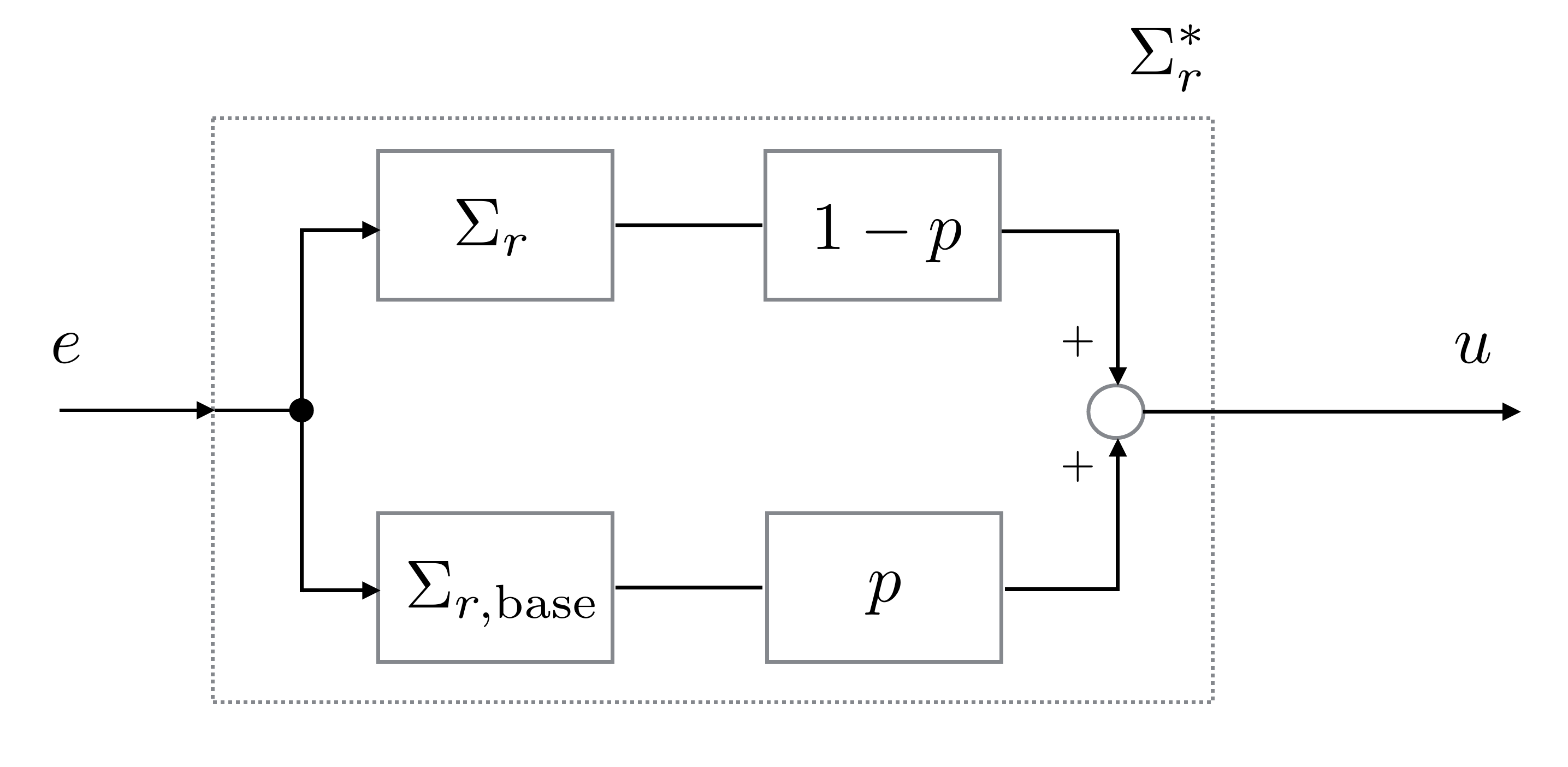}
\caption{}
\label{fig:convex}
\end{subfigure}
\caption{Block diagram of the (\subref{fig:PI+CI}) PI+CI controller, adapted from \cite{banos2011reset} and (\subref{fig:convex}) convex combination of the reset controller $\Sigma_r$ and its linear base controller $\Sigma_{r,\mathrm{base}}$.}
\end{figure}

\begin{figure*}[!htb]
\centering
\begin{subfigure}{0.325\linewidth}
\flushleft
\includegraphics[width=\linewidth]{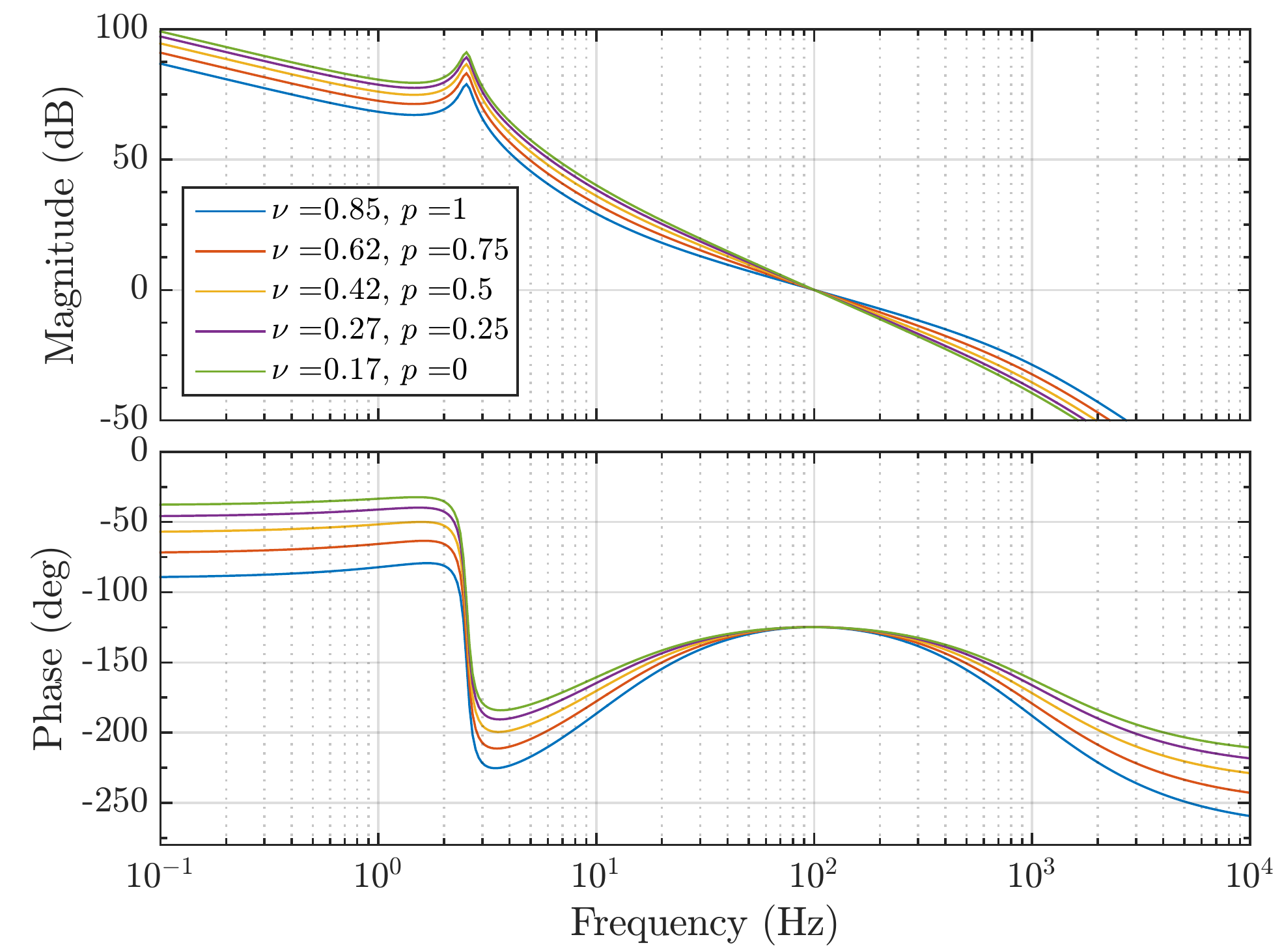}
\subcaption{}
\label{fig:picibode}
\end{subfigure}
\begin{subfigure}{0.325\linewidth}
\centering
\includegraphics[width=\linewidth]{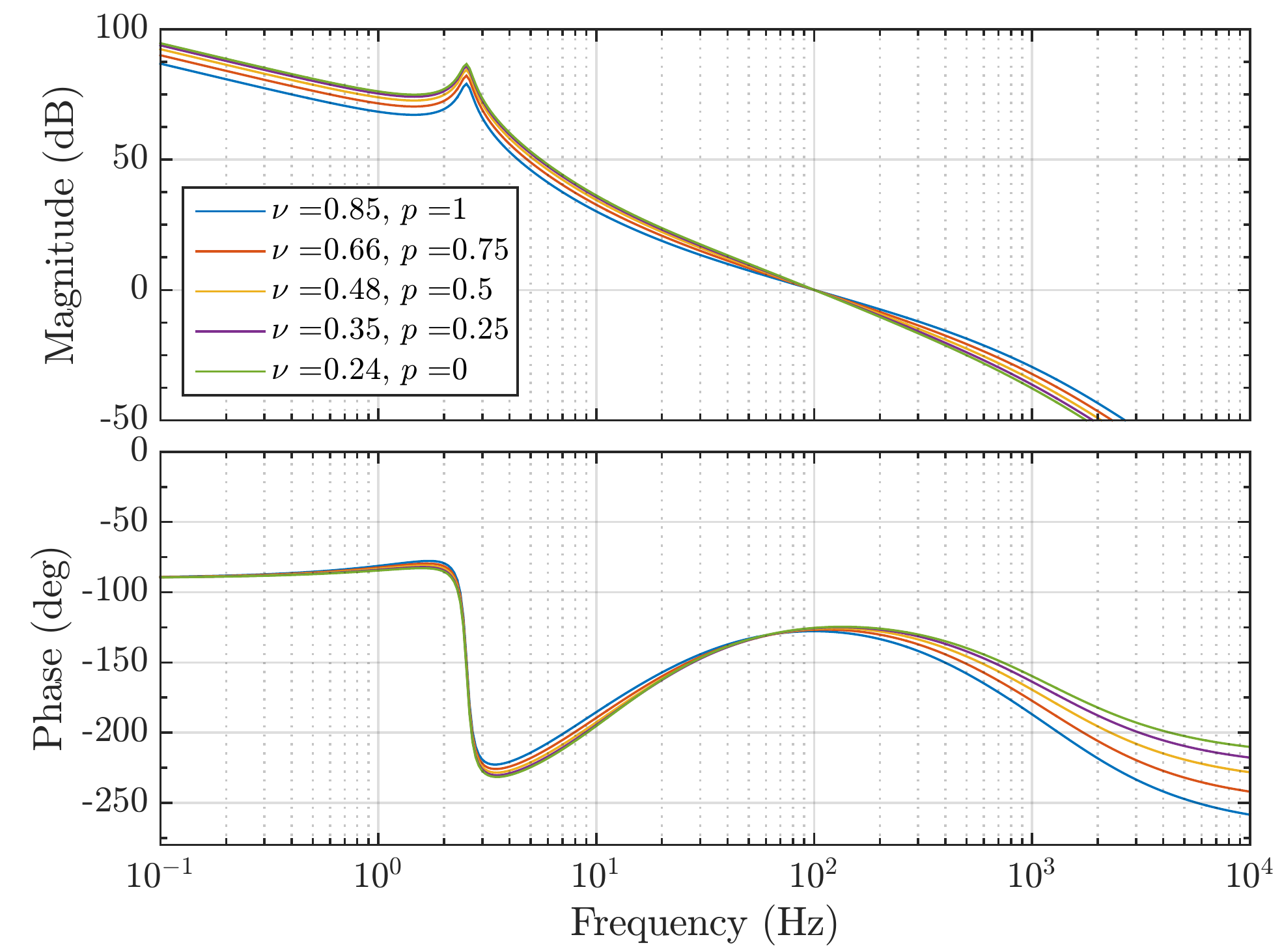}
\subcaption{}
\label{fig:lagbode}
\end{subfigure}
\begin{subfigure}{0.325\linewidth}
\flushright
\includegraphics[width=\linewidth]{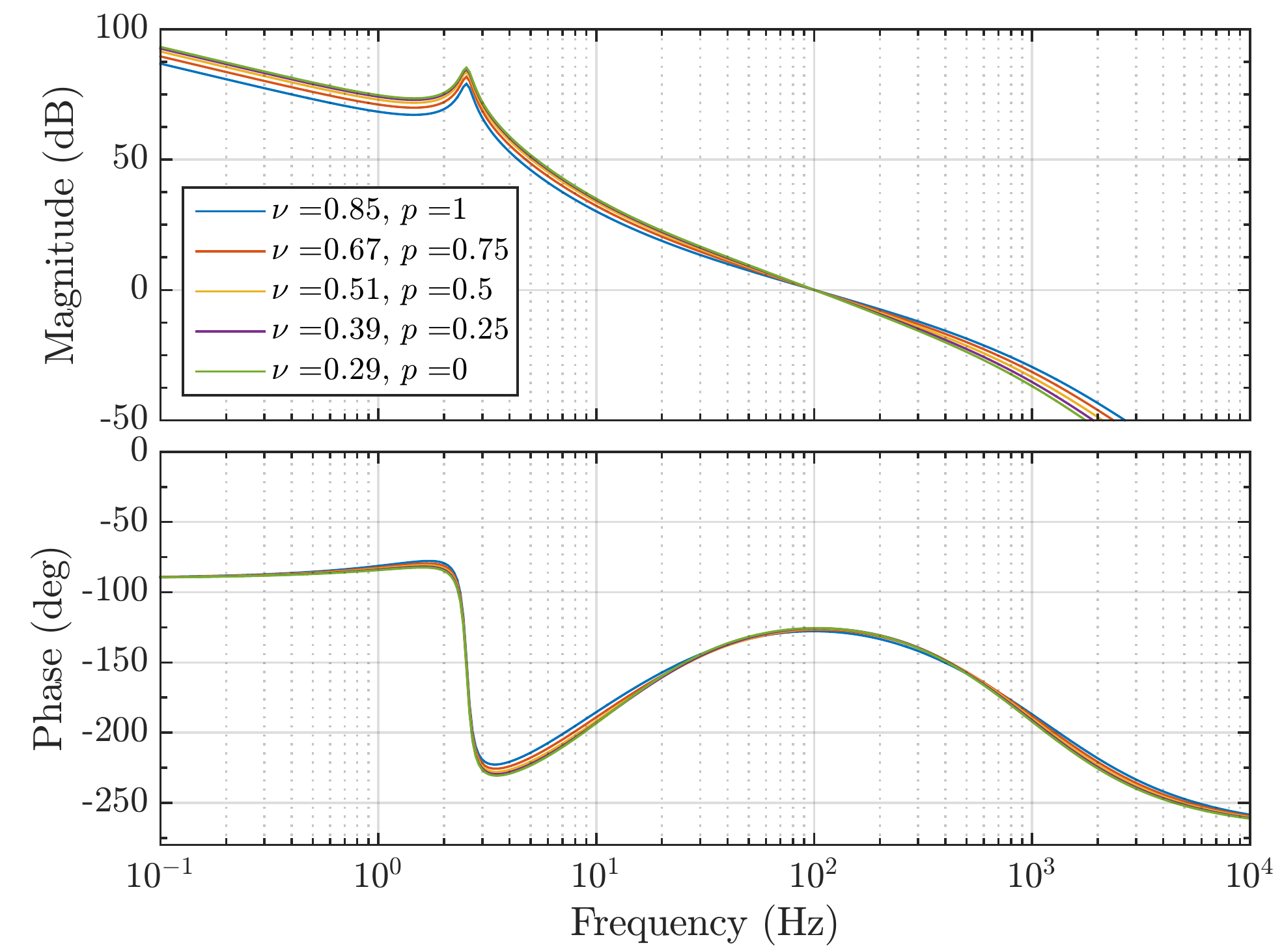}
\subcaption{}
\label{fig:leadlagbode}
\end{subfigure}
\caption{Open-loop for first generation CRONE with second order plant as in (\ref{eq:plant}) without delay for $\gamma=0$ and different $p$-values with (\subref{fig:picibode}) integrator reset, (\subref{fig:lagbode}) first order filter reset and (\subref{fig:leadlagbode}) lag reset. For the same phase margin at similar bandwidth the open-loop gain behaviour with reset is favourable.}
\label{fig:bode}
\end{figure*}

\subsection{Control structure}\label{subsec:struct}
The proposed CRONE reset control design structure firstly includes the choice of which part of the CRONE transfer function will be reset. In addition a generalized structure is proposed that allows for tuning the amount of non-linearity versus linearity of the system.
The transfer functions below are given for first generation CRONE. 
\subsubsection{CRONE reset strategies}\label{subsec:strategy}
\paragraph{CRONE integrator reset}

The CRONE integrator reset strategy can be represented by following equation:

\begin{flalign}
&C_\mathrm{int}(s)=\underbrace{(\frac{\omega_I}{s} )^{n_I-1}C_0(\frac{s}{\omega_I}+1)^{n_I}\bigg(\frac{1+\frac{s}{\omega_b}}{1+\frac{s}{\omega_h}}\bigg)^\nu\frac{1}{(1+\frac{s}{\omega_F})^{n_F}}}_{\Sigma_{nr}}\nonumber\\
&\qquad\qquad\qquad\qquad\qquad\qquad\qquad\qquad\qquad\quad \times\underbrace{\frac{\omega_I}{s}}_{\Sigma_r}
\end{flalign}

in which the integrator part becomes the part that is reset $\Sigma_r$ and the rest of the transfer function is linear $\Sigma_{nr}$.


\paragraph{CRONE first order filter reset}
Similarly, CRONE first order filter reset strategy can be described as:
\begin{equation}
C_\mathrm{fof}(s)=\underbrace{\frac{(1+\frac{s}{\omega_h})^\nu}{(1+\frac{s}{\omega_h})^{\nu+1}} C_0(1+\frac{\omega_I}{s})^{n_I}\frac{1}{(1+\frac{s}{\omega_F})^{n_F}}}_{\Sigma_{nr}}\underbrace{\frac{1}{1+\frac{s}{\omega_b}}}_{\Sigma_r}.
\end{equation}

\paragraph{CRONE lag reset}
Finally, the CRONE lag reset strategy can be described as:
\begin{equation}
C_\mathrm{lag}(s)=\underbrace{\bigg(\frac{1+\frac{s}{\omega_b}}{1+\frac{s}{\omega_h}}\bigg)^{\nu+1} C_0(1+\frac{\omega_I}{s})^{n_I}\frac{1}{(1+\frac{s}{\omega_F})^{n_F}}}_{\Sigma_{nr}}\underbrace{\frac{1+\frac{s}{\omega_h}}{1+\frac{s}{\omega_b}}}_{\Sigma_r}.
\end{equation}


\subsubsection{Two degree of freedom non-linearity tuning controller}\label{subsec:convex}
As the CRONE reset controller is designed using describing function, higher harmonics introduced by non-linearity are neglected. However these harmonics do affect the performance of the system. Hence, tuning parameters are needed to control the level of non-linearity introduced in the system. Two different tuning parameters are considered in this work, which will be explained below. Then both methods are used in the proposed control structure.

\begin{figure}[!htb]
\centering
\begin{subfigure}{0.47\linewidth}
\includegraphics[width=\linewidth]{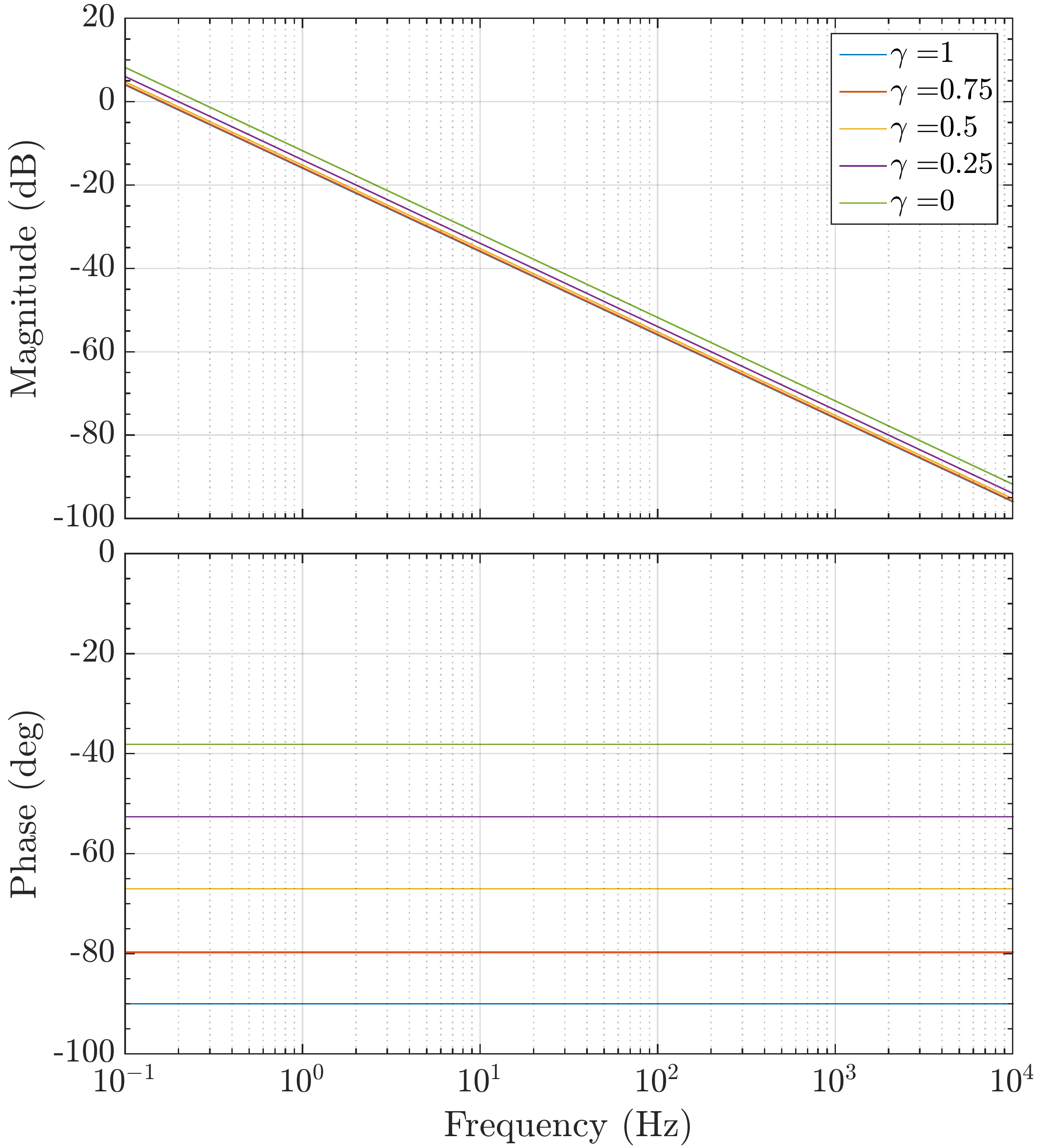}
\caption{}
\label{fig:DFgamma}
\end{subfigure}
\begin{subfigure}{0.47\linewidth}
\includegraphics[width=\linewidth]{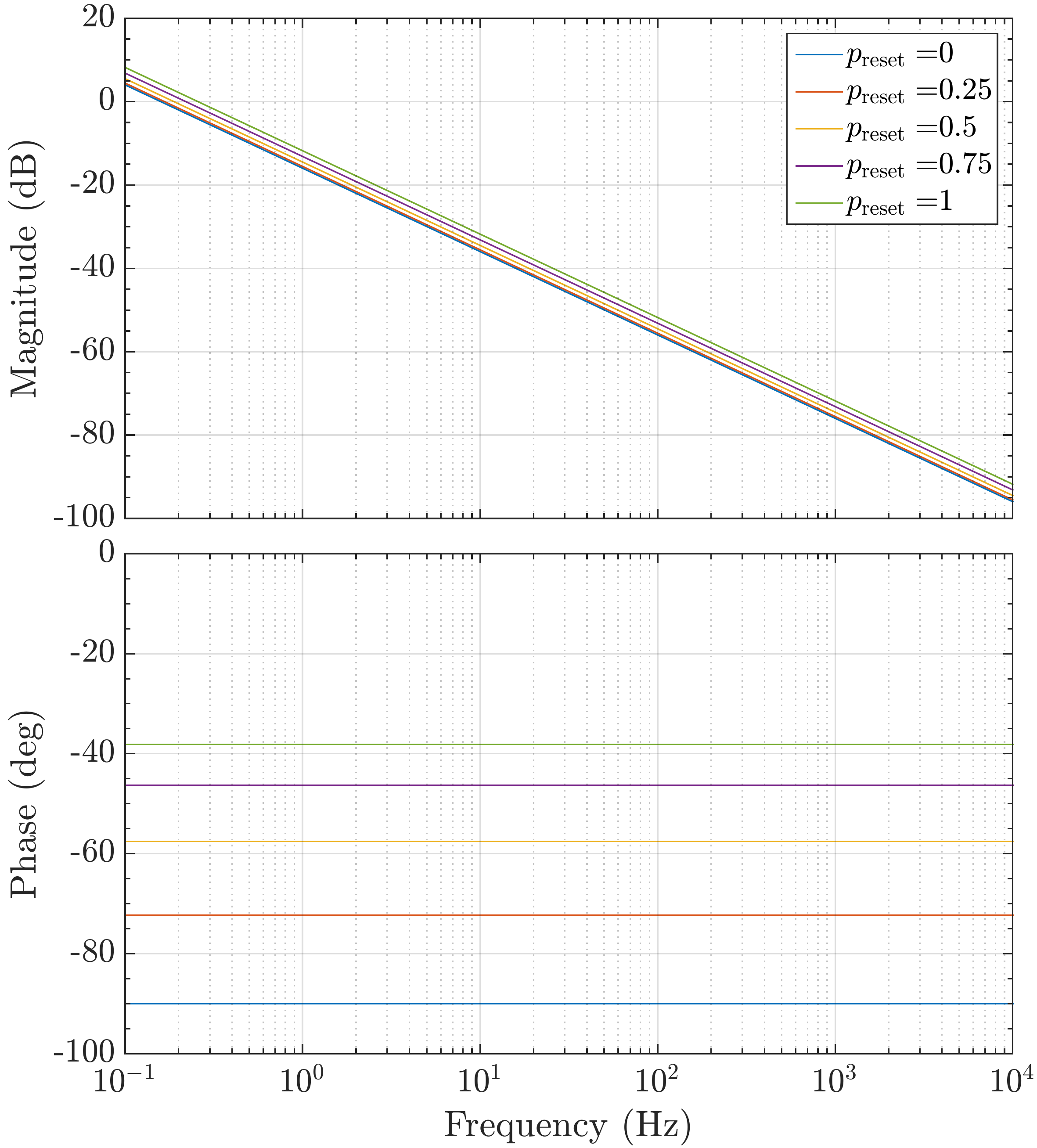}
\caption{}
\label{fig:DFpreset}
\end{subfigure}
\caption{Describing function of a reset integrator when tuning non-linearity using variable (\subref{fig:DFgamma}) $\gamma$ and using variable (\subref{fig:DFpreset}) $p_\mathrm{reset}$.}
\label{fig:DFtuning}
\end{figure}

\paragraph{Reset fraction $\gamma$}
In the first method the reset matrix $A_\rho$ can be taken as $A_\rho=\gamma I$ \cite{banos2011reset}. Then the state can be reset to a fraction $\gamma$. With $\gamma=1$, $\Sigma_r$ becomes linear, whereas with $\gamma=0$ a full reset occurs. The describing function of a reset integrator is shown in Fig. \ref{fig:DFgamma}, and it becomes evident that the reset phase lead can be tuned with $\gamma$.

\paragraph{Reset percentage $p_\mathrm{reset}$}
The second way of tuning non-linearity in a reset system is using a PI+CI approach in the same work of \cite{banos2011reset}, including a reset percentage $p_\mathrm{reset}$. This results in a convex combination of a linear integrator and a Clegg integrator. A block diagram of a PI+CI controller can be found in Fig. \ref{fig:PI+CI}. The describing function of a reset integrator with varying $p_\mathrm{reset}$ is shown in Fig. \ref{fig:DFpreset}. Intuitively, the amount of phase lead achieved with reset is less when less non-linearity is allowed in the system.\\



The parallel interconnected system of $\Sigma_r$ and $\Sigma_{r,\mathrm{base}}$ in {Fig. \ref{fig:convex}} has a more general approach compared to PI+CI : other filters than integrator can be chosen to be combined. Following impulsive differential equations represent this system structure for $\xi=\begin{bmatrix}
x_{r}^T, x_{r,\mathrm{base}}^T
\end{bmatrix}^T$:
\begin{eqnarray}
\Sigma^*_r:&\begin{cases}
\dot{\xi}=\underbrace{\begin{bmatrix}
A_r&0\\0&A_r
\end{bmatrix}}_{\bar{A}}\xi+\underbrace{\begin{bmatrix}
B_r\\B_r
\end{bmatrix}}_{\bar{B}}e,& \text{if }  e\neq 0,\\
\xi(t^+)=\underbrace{\begin{bmatrix}
\gamma I&0\\0&I
\end{bmatrix}}_{\bar{A_\rho}}\xi,& \text{if } e=0,\\
u_r=\underbrace{\begin{bmatrix}
(1-p)C_r&pC_r
\end{bmatrix}}_{\bar{C}}\xi+\underbrace{D_r}_{\bar{D}}e.
\end{cases}
\end{eqnarray}
$\Sigma^*_r$, depicted in Fig. \ref{fig:convex} is used instead of $\Sigma_r$ in the control diagram of Fig. \ref{fig:rcontrol}. When taking $A_r=\bar{A}$, $B_r=\bar{B}$, $C_r=\bar{C}$, $D_r=\bar{D}$, $A_\rho=\bar{A_\rho}$ and taking (\ref{eq:DF}), the describing function of this controller part can be found. Note that $p$ is taken as $1-p_\mathrm{reset}$ such that with $p=1$ or $\gamma=1$ the system is a fully linear system, and full reset if $p=0$ and $\gamma=0$.

\subsection{Renewed calculation of fractional order $\nu$}

The additional phase at bandwidth is given by $\Phi_r(\omega_{cg})$, which is calculated from:
\begin{equation}
\Phi_r(\omega)=\angle G_\mathrm{DF}(j\omega)-\angle G(j\omega)
\end{equation}
in which $G_\mathrm{DF}(j\omega)$ is the describing function of the reset filter and $G(j\omega)$ is the transfer function of its linear base system. Then the new fractional order for first generation CRONE reset $\nu^*$ can be calculated as:

\begin{multline}
\nu^*=\frac{-\pi+M_\Phi-\arg G(j\omega_{cg})+n_F\arctan\frac{\omega_{cg}}{\omega_F}}{\arctan\frac{\omega_{cg}}{\omega_b}-\arctan\frac{\omega_{cg}}{\omega_h}}\\+\frac{n_I(\frac{\pi}{2}-\arctan \frac{\omega_{cg}}{\omega_I})-\Phi_r}{\arctan\frac{\omega_{cg}}{\omega_b}-\arctan\frac{\omega_{cg}}{\omega_h}}.
\end{multline}


The amount of phase advantage calculated for different reset strategies in first generation CRONE are as follows:

\begin{enumerate}[label=\alph*)]
\item \textit{CRONE integrator reset}\\
\begin{equation}\label{eq:phiint}
\Phi_{r,\mathrm{int}}(p,\gamma)=\arctan \bigg(\frac{4}{\pi}(1-p)\frac{1-\gamma}{1+\gamma}\bigg)
\end{equation}
which is a constant phase for all frequencies when fixing $p$ and $\gamma$.\\
\item  \textit{CRONE first order filter reset}\\
\begin{equation}\label{eq:philag}
\Phi_{r,\mathrm{fof}}(\omega,p,\gamma,\omega_b)=\arctan\bigg((1-p)\Theta_D(\gamma,\omega,\omega_b)\bigg)
\end{equation}\\
with $\Theta_D(\gamma,\omega)$ defined as (\ref{eq:thetad}).

\item \textit{CRONE lag reset}\\
\begin{multline}\label{eq:phill}
\Phi_{r,\mathrm{lag}}(\omega, p, \gamma, \omega_b, \omega_h)=\\ \arctan\bigg(\frac{(1-p)\Theta_D(\gamma,\omega,\omega_b)(1-\frac{\omega_b}{\omega_h})}{1+(\frac{\omega}{\omega_h})^2+\frac{\omega}{\omega_h}\Theta_D(\gamma,\omega,\omega_b)(1-\frac{\omega_b}{\omega_h})}\bigg)
\end{multline}
\end{enumerate}
%


The open-loop responses with these reset strategies can be found in Fig. \ref{fig:bode}. In all of the strategies the open-loop shape is better with increasing non-linearity (smaller $p$ value). For full reset ($p=0$ and $\gamma=0$) the gain advantage with respect to the linear CRONE controller at 1\si{\kilo\hertz} is 10.9\si{\deci\bel}, 8.2\si{\deci\bel} and 7.4\si{\deci\bel}, for integrator reset, first order filter reset and lag reset respectively with bandwidth at 100\si{\hertz}.

\FloatBarrier

\section{Stability}\label{sec:stability}
%

The closed-loop system depicted in {Fig. \ref{fig:rcontrol}} is a reset system for which the following stability result holds: 

\begin{thm}{\cite{banos2011reset}.}
Let $V: \mathbb{R}^n\rightarrow\mathbb{R}^n$ be a continuously differentiable, positive-definite, radially unbounded function such that
\begin{eqnarray}\label{eq:lyapunov}
\dot{V}(x):=\Big(\pdv{V}{x}\Big)^TA_{cl}x<0,&\text{if } e(t) \neq 0,\\
\Delta V(x):=V(A_\rho x)-V(x)\leq 0,& \text{if } \label{eq:lyapunov2} e(t)=0
\end{eqnarray}
where $A_{cl}$ is the closed-loop $A$-matrix:
\begin{equation}
A_{cl}=\begin{bmatrix}
\bar{A}&\bar{B}C_{nrp}\\
-B_{nrp}\bar{C}&A_{nrp}
\end{bmatrix}
\end{equation}
where $(\bar{A},\bar{B},\bar{C},\bar{D})$ are the state-space matrices of $\Sigma^*_r$ and $(A_{nrp}, B_{nrp},C_{nrp}, D_{nrp})$ are the state-space matrices of non-reset controller $\Sigma_{nr}$ and plant $\Sigma_p$ combined in series, and $A_\rho$ is defined as:
\begin{equation}
A_\rho=\mathrm{diag}(\bar{A}_\rho,I_{nrp})
\end{equation}
where $\bar{A}_\rho$ is the reset matrix of $\Sigma^*_r$ and $n_{nrp}$ is sum of the number of states of non-reset controller $n_{nr}$ and plant $n_p$. 

Then the reset control system is asymptotically stable.
\end{thm}

Quadratic stability is guaranteed when (\ref{eq:lyapunov}) and (\ref{eq:lyapunov2}) hold true for a potential function $V(x)=x^TPx$ with $P>0$. From this condition the authors of \cite{banos2011reset} obtained following theorem for proving quadratic stability:

\begin{thm}{\cite{banos2011reset}.} 
There exists a constant $\beta\in\mathbb{R}^{n_r\times 1}$ and $P_\rho\in\mathbb{R}^{n_r\times n_r}, P_\rho>0$ where $n_r$ is the number of reset states, such that the restricted Lyapunov equation
\begin{eqnarray}\label{eq:lmi1}
P>0,&A_{cl}^TP+PA_{cl}<0,\\
&B_0^TP=C_0
\end{eqnarray}
has a solution for P, where $C_0$ and $B_0$ are defined by:
\begin{eqnarray}
C_0=\begin{bmatrix}
\beta C_{nrp}&O_{n_r\times n_{rnr}}&P_\rho
\end{bmatrix},
B_0=\begin{bmatrix}
O_{n_{nrp}\times n_r}\\O_{n_{rnr}\times n_r}\\I_{n_r}
\end{bmatrix}\label{eq:lmi2}
\end{eqnarray}
and $n_{rnr}$ is the number of non-reset states in $\Sigma^*_r$.
\end{thm}

\section{Experimental validation}\label{sec:validation}
\subsection{Experimental setup}
\begin{figure}[!htb]
\centering
\begin{subfigure}{\linewidth}
\centering
\includegraphics[width=.75\linewidth]{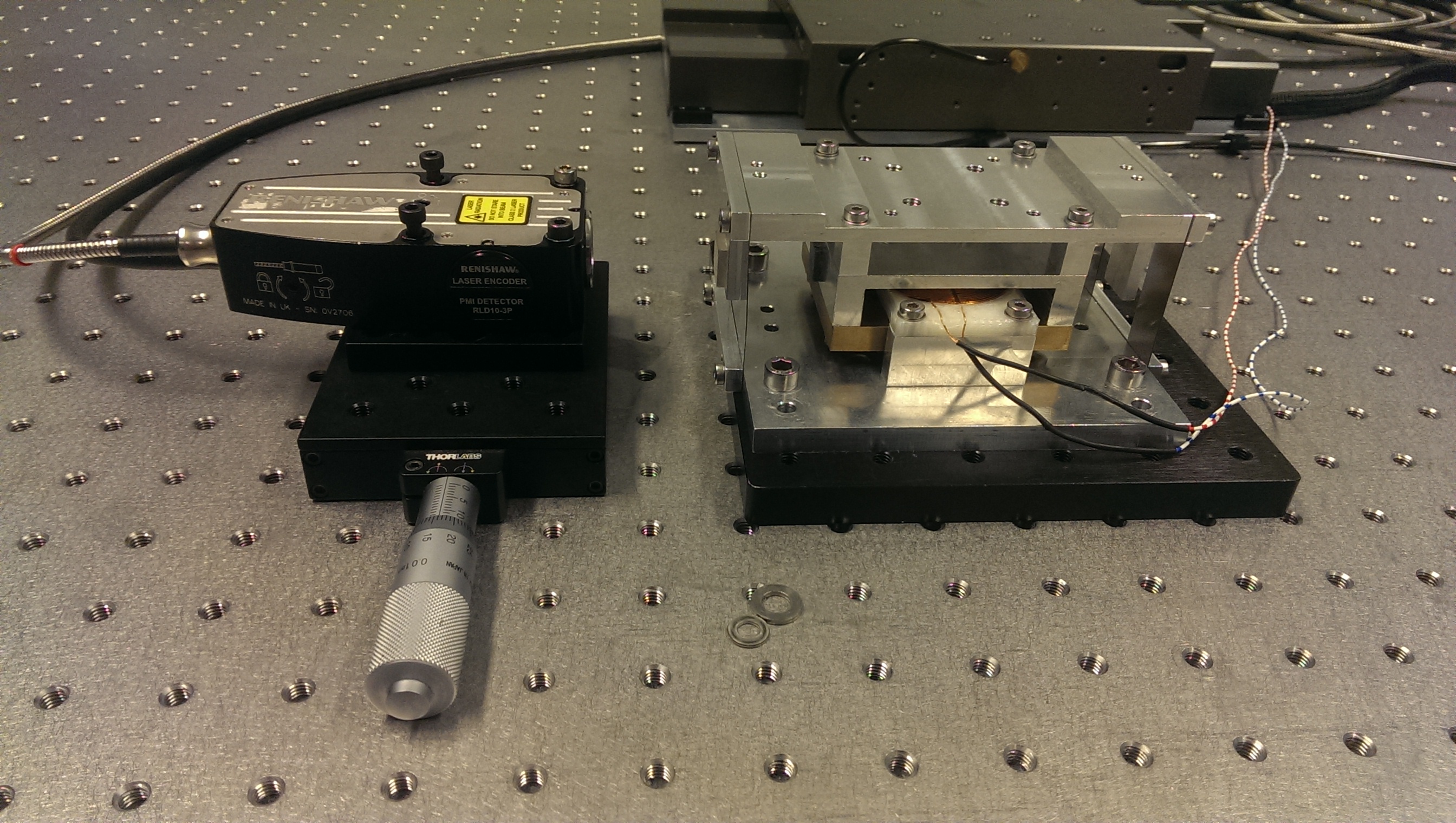}
\caption{}
\label{fig:setup}
\end{subfigure}\\
\begin{subfigure}{\linewidth}
\centering
\includegraphics[width=.85\linewidth]{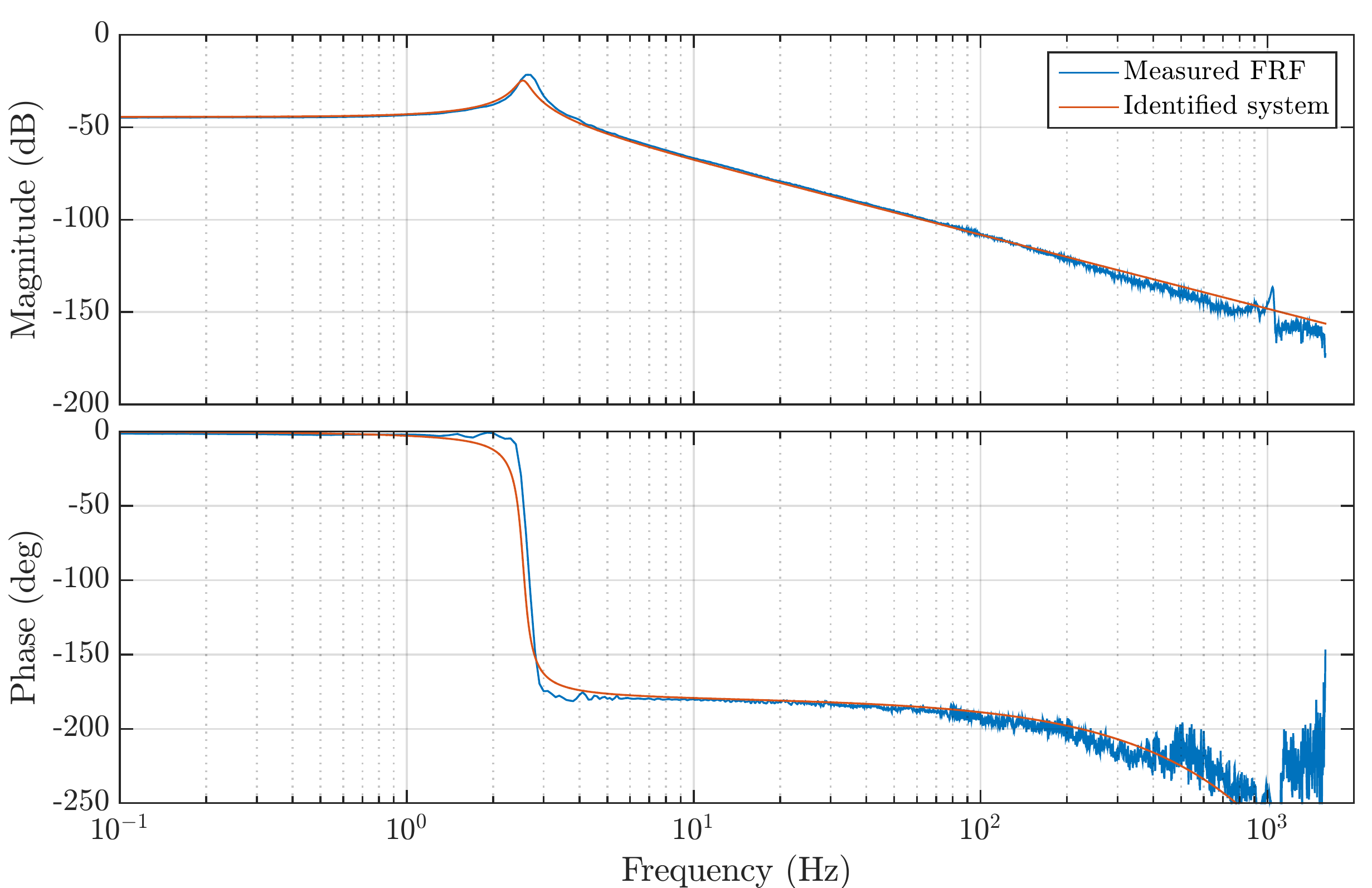}
\caption{}
\label{fig:frf}
\end{subfigure}
\caption{(\subref{fig:setup}) Picture of the Lorentz stage (right) with the laser encoder at the left. (\subref{fig:frf}) Frequency response of the system and the identified system 
model.}
\end{figure}

The practical system used for validation of the CRONE reset control design is a custom-designed precision stage that is actuated with the use of a Lorentz actuator. This stage is linear-guided using flexures to attach the Lorentz actuator to the base of the stage and actuated at the center of the flexures. With a \textit{Renishaw RLE10} laser encoder the position of the fine stage is read out with 10\si{\nano\metre} resolution. A picture of the setup can be found in Fig. \ref{fig:setup}. The controller is designed within a MATLAB/Simulink environment and implemented digitally via dSPACE DS1103 real-time control software. A sampling rate of 20\si{\kilo\hertz} is used for all implemented controllers.

The frequency response of the system is shown in Fig. \ref{fig:frf} and shows the behaviour of a second order mass-spring-damper system with additional harmonics at higher frequencies and phase lag due to delay. The transfer function of this system is identified as:

\begin{equation}\label{eq:plant}
P(s)=\frac{0.5474}{0.5718s^2+0.95+146.3}e^{\num{-2.5e-4}s} .
\end{equation}

%

\subsection{Stability analysis}
In this subsection the stability theory presented in previous section is applied to the designed CRONE reset controllers.
As a practical example the stability analysis is shown for a CRONE-1 lag reset controller with fourth-order approximation of the fractional order (using Oustaloup approximation as in \cite{sabatier2015fractional}), controller structure as described in section \ref{subsec:struct} and parameters as summarized in Table \ref{tab:parameters}. 

The non-reset control and reset control parts of the CRONE reset controller are then given by (\ref{eq:cronereset2}) and (\ref{eq:cronereset1}) respectively.

\begin{equation}\label{eq:cronereset1}
C_{r}(s)=\frac{78.54s+\num{3.948e5}}{5027s+\num{3.948e5}}
\end{equation}

\begin{figure*}[!htb]
\begin{equation}\label{eq:cronereset2}
C_{nr}(s)=\frac{\num{2.387e10} s^5 + \num{4.589e13} s^4 + \num{2.351e16} s^3 + \num{4.084e18 }s^2 + \num{2.618e20} s  + \num{5.553e21}}{0.0285 s^6 + 530.6 s^5 + \num{3.267e6} s^4 + \num{7.426e9} s^3 + \num{5.722e12} s^2+ \num{1.175e15}s}                                                                     
\end{equation}
\end{figure*}

The LMI consisting of (\ref{eq:lmi1}) to (\ref{eq:lmi2}) for this controller and identified plant is solved using a Sedumi solver from the YALMIP toolbox in MATLAB \cite{Lofberg2004}, and gives a solution for $\beta=-\num{8.99e-12}$ and $P_\rho= 0.98$. For the found $P$-matrix (after normalizing $A_{cl}$), the eigenvalues are $\num{2.64e-11}$, $\num{1.86e-3}$, $0.98$, $1.00$, $1.00$, $1.00$, $1.00$, $1.00$, $1.00$ and $2.00$. Because the eigenvalues are all larger than zero, the found $P$ is indeed positive definite. Thus quadratic stability is guaranteed.

\begin{table}
\centering
\caption{Parameter values for CRONE-1 lag controller}
\begin{tabular}{lll}\toprule
Symbol&Parameter&Value\\\midrule
PM&phase margin&55\si{\degree}\\
$\omega_{cg}$&bandwidth&100\si{\hertz}\\
$\omega_b$&lead corner frequency&12.5\si{\hertz}\\
$\omega_h$&lag corner frequency&800 \si{\hertz}\\
$\omega_I$&integrator corner frequency&8.33\si{\hertz}\\
$\omega_F$&low-pass filter corner frequency&1200\si{\hertz}\\
$n_I$&integrator order&1\\
$n_F$&low-pass filter order&1\\
$p$&reset percentage&0.5\\
$\gamma$&partial reset&0.5\\\bottomrule
\end{tabular}
\label{tab:parameters}
\end{table}
 
\section{Results}\label{sec:results}
The CRONE reset controller is designed using describing function. Since reset control is non-linear, it is important to verify that the expected frequency behaviour matches actual response. For this purpose, practical frequency domain results have been retrieved.  Firstly, sensitivity- and complementary sensitivity function results are discussed which are used to compute an open-loop response. Then reference-tracking results are shown for tracking an input-shaped triangular wave. Finally, time domain noise attenuation results are analysed. Since most existing work in literature is done on integrator reset and for reasons of brevity the results below are given for CRONE-1 lag reset. 
\subsection{Identification sensitivity $S(j\omega)$  and complementary sensitivity $T(j\omega)$ }

\begin{figure}[H]
\centering
\includegraphics[width=.95\linewidth]{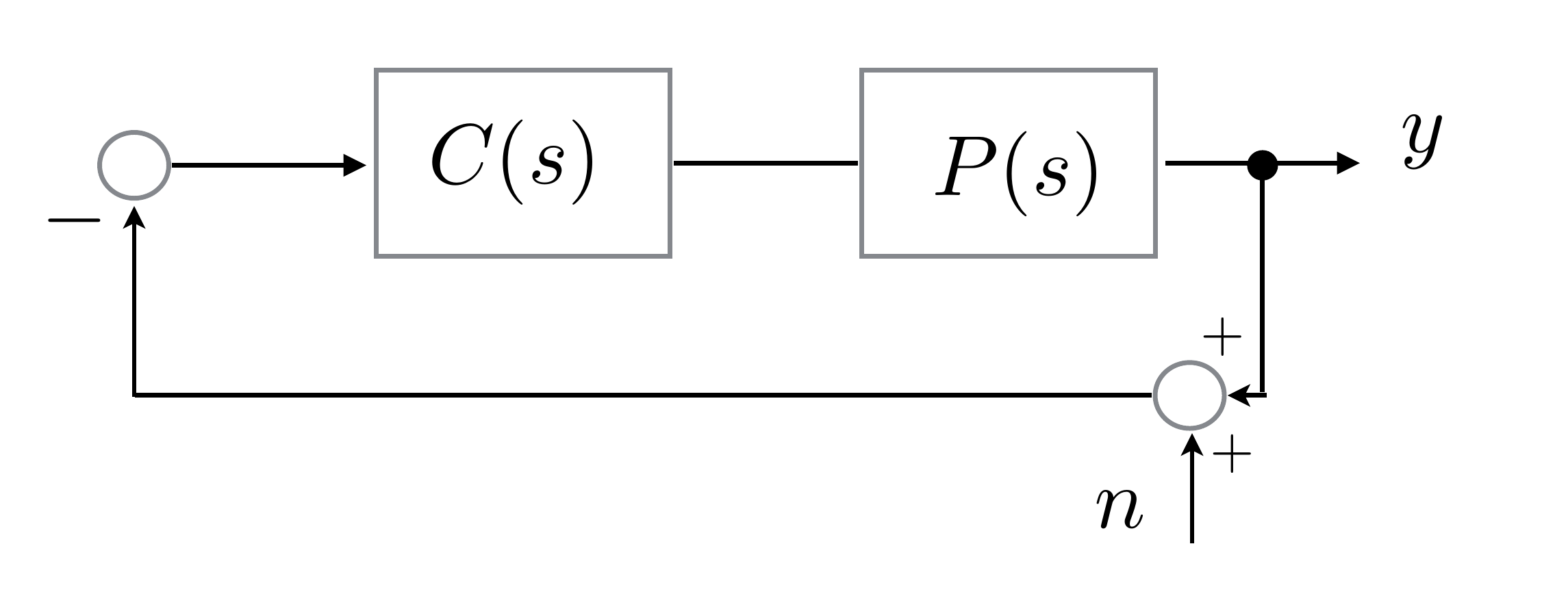}
\caption{Block diagram of the control loop and signals used for identification of $T(j\omega)$ and $S(j\omega)$.}
\label{fig:blockdiagsens}
\end{figure}

\begin{figure}[H]
\centering
\begin{subfigure}{.95\linewidth}
\includegraphics[width=\linewidth]{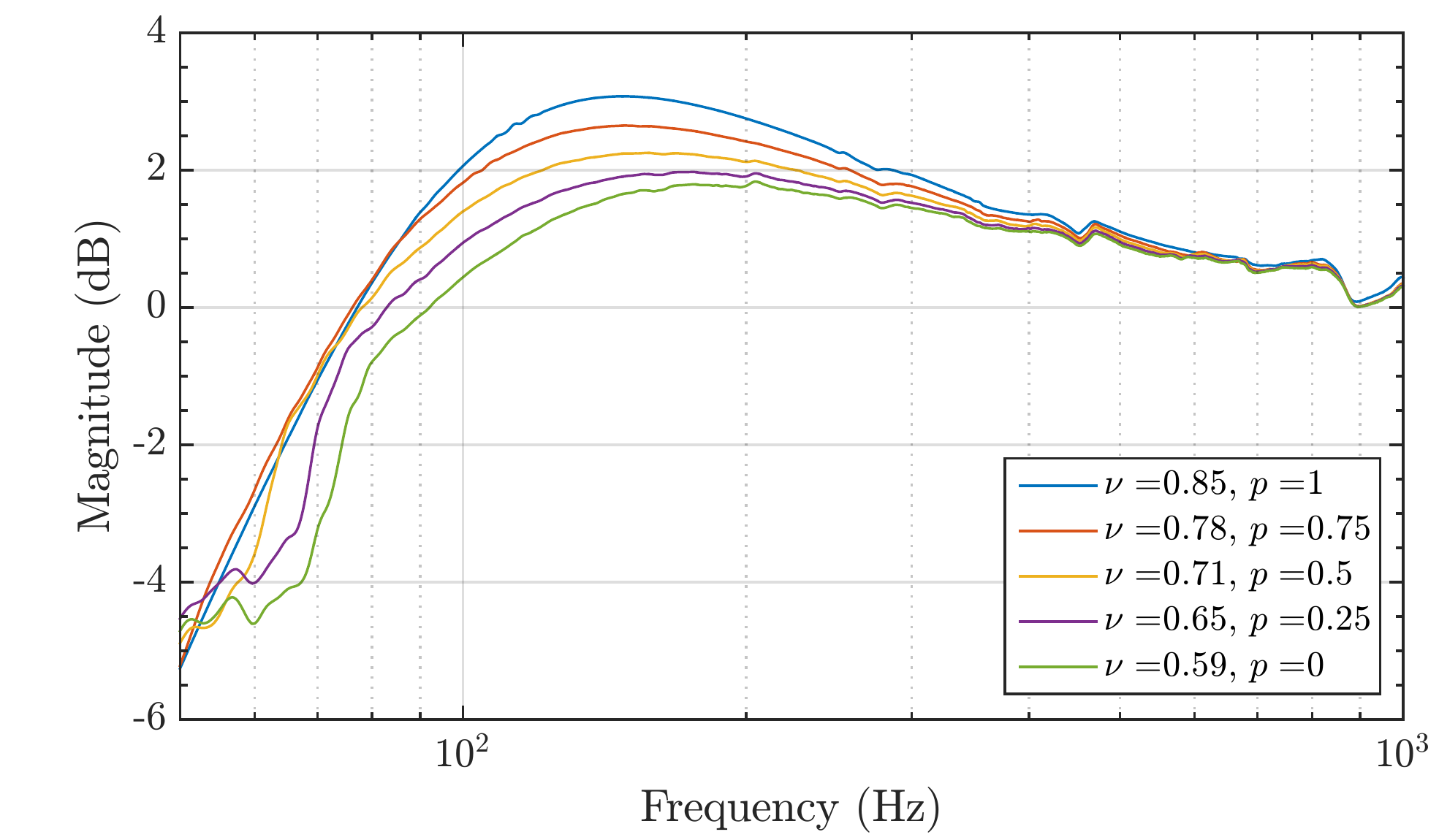}
\caption{}
\label{fig:sens}
\end{subfigure}\\
\begin{subfigure}{.95\linewidth}
\includegraphics[width=\linewidth]{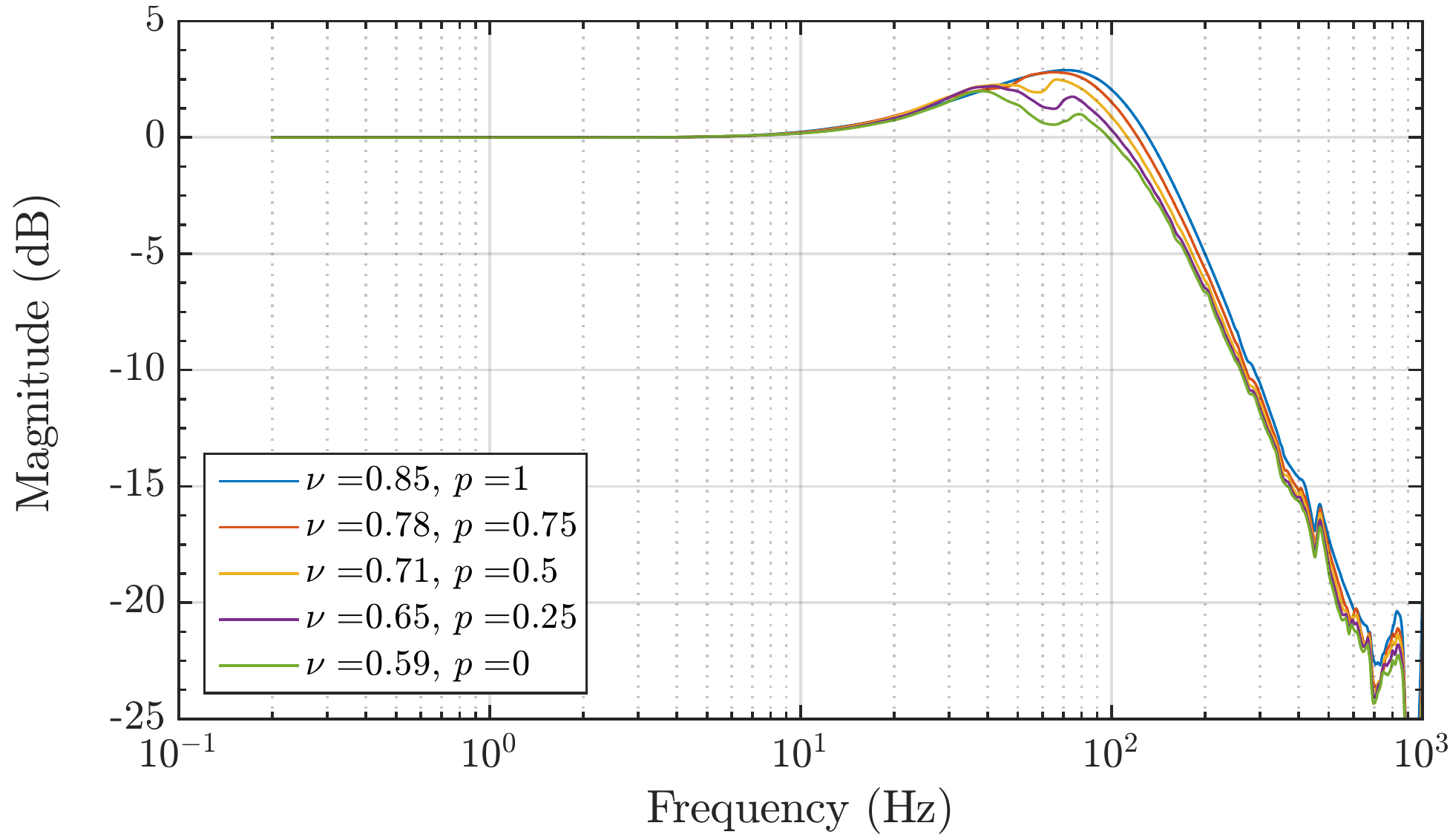}
\caption{}
\label{fig:csens}
\end{subfigure}
\caption{(\subref{fig:sens}) Sensitivity and (\subref{fig:csens}) complementary sensitivity function for CRONE-1 lag reset with constant $\gamma=0.5$ and different $p$-values.}
\label{fig:senss}
\end{figure}
Both  $T(j\omega)$ and $S(j\omega)$ were identified using a frequency sweep at the position of signal $n$ in {Fig. \ref{fig:blockdiagsens}}. $T(j\omega)$ then can be identified as the transfer from $-n$ to $y$, whereas $S(j\omega)$ is identified as the transfer from $n$ to $y+n$. The frequency sweep was done for frequencies 0.1\si{\hertz} to 2.5\si{\kilo\hertz} with a target time of 120\si{\second} and a total duration of 480\si{\second}.

$S(j\omega)$ can be found in {Fig. \ref{fig:sens}}. $T(j\omega)$ is shown in Fig. \ref{fig:csens}.  As expected, noise transfer at high frequencies is reduced for lower $\nu$ and this is seen in both figures. 
However, for linear controllers, this would have resulted in an increase in sensitivity peak due to waterbed effect. But it is seen in {Fig. \ref{fig:sens}} that with reset control, this effect is countered with the sensitivity peak below that of linear case ($p = 1$). This is also evident with complementary sensitivity in {Fig. \ref{fig:csens}} where we additionally see larger decrease in gain at higher frequencies compared to linear control, hence overcoming waterbed effect.

\subsection{Open-loop response}
The open-loop responses for different CRONE-1 reset controllers were estimated by a point-wise division of identified complementary sensitivity function $T(j\omega)$ and sensitivity function $S(j\omega)$ for the same range of frequencies. This response is analysed to validate that phase margin is satisfied in all CRONE reset controllers. The identified open-loop at low frequencies up till 10\si{\hertz} is considered unreliable as the coherence between the signals used for identification of the sensitivity function is poor. 

\begin{figure}[!htb]
\centering
\begin{subfigure}{.47\linewidth}
\includegraphics[width=\linewidth]{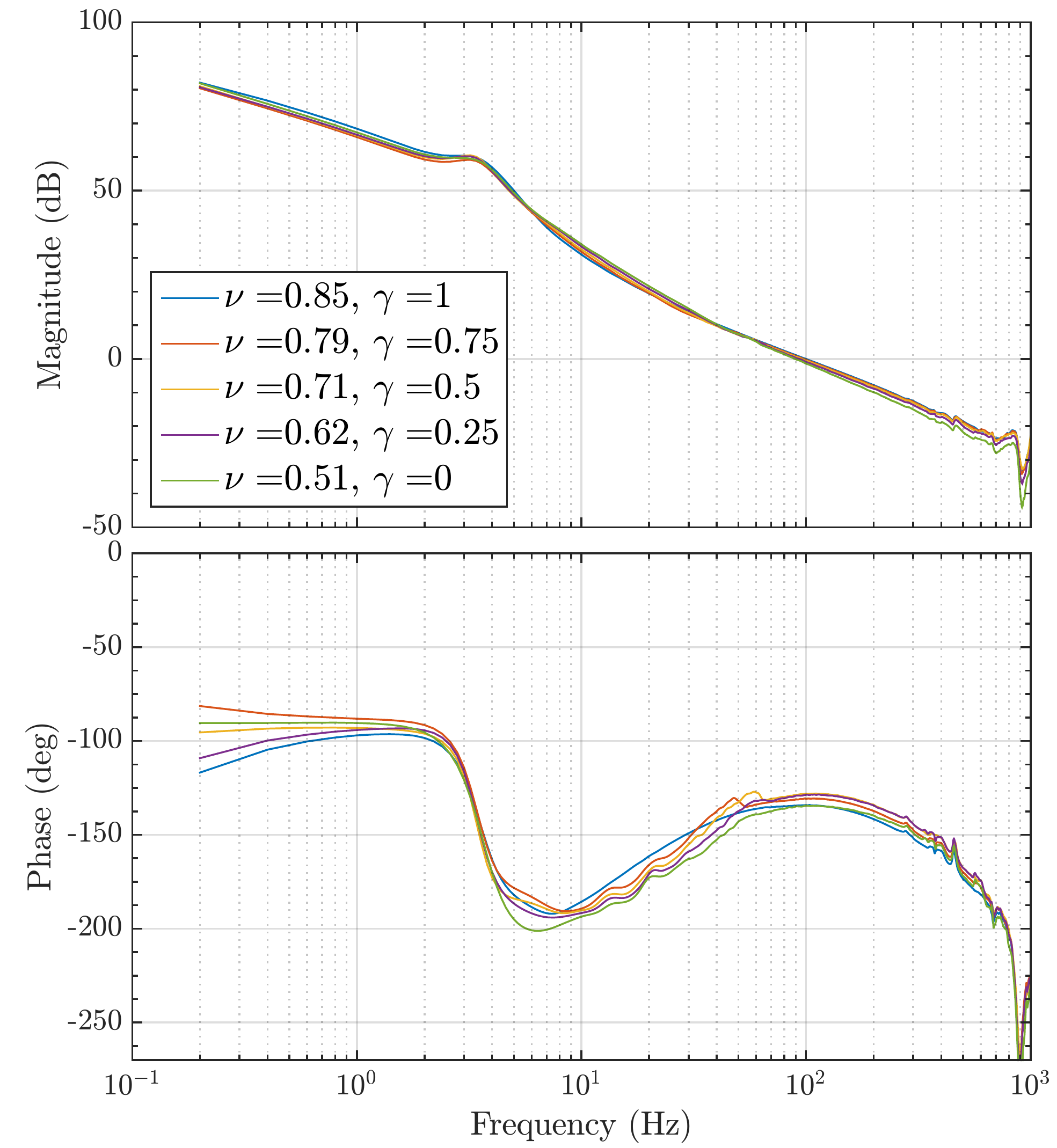}
\caption{}
\label{fig:llmeas}
\end{subfigure}
\begin{subfigure}{.47\linewidth}
\includegraphics[width=\linewidth]{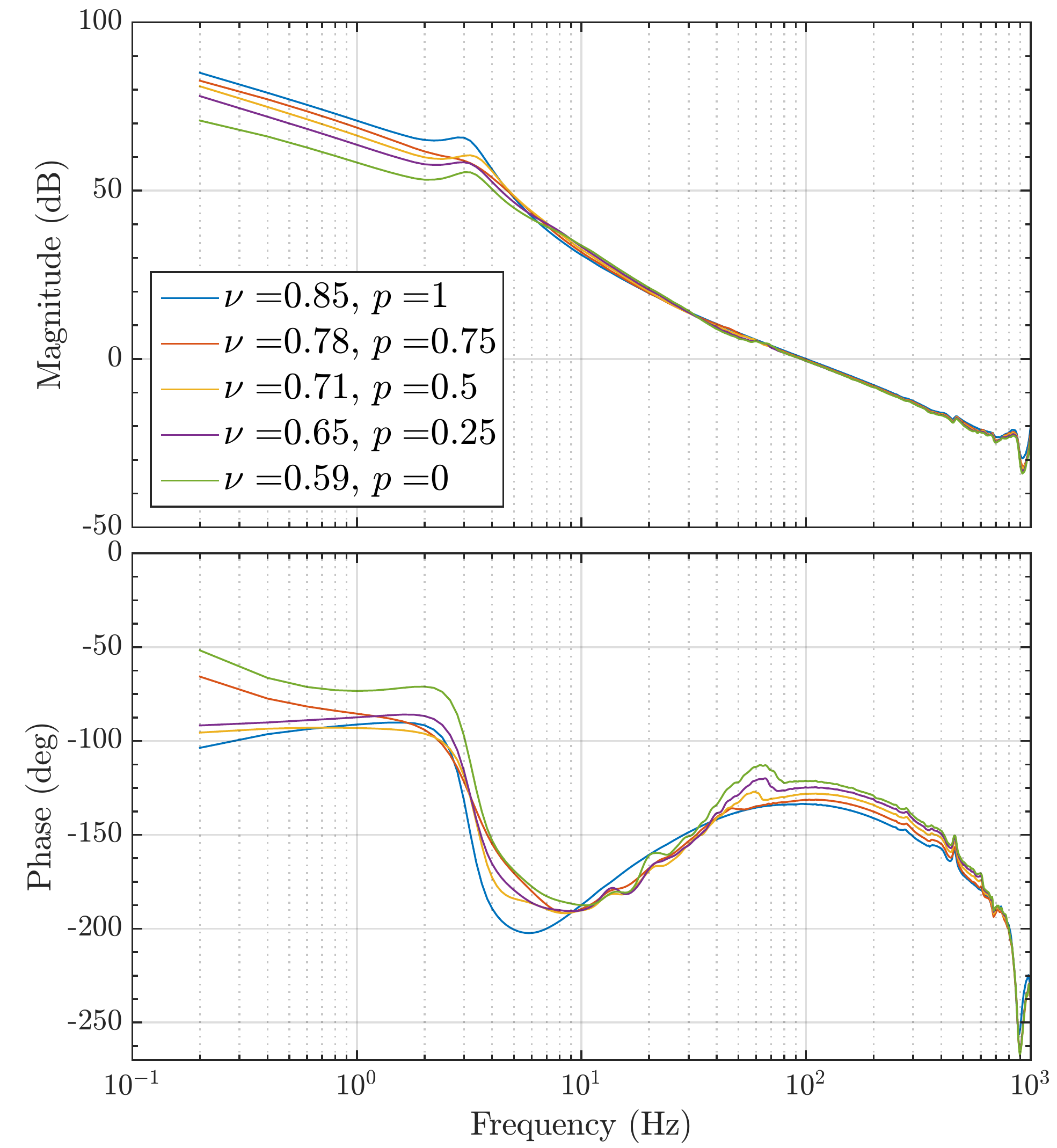}
\caption{}
\label{fig:llmeasg}
\end{subfigure}
\caption{Experimental open-loop responses for the system with CRONE-1 (\subref{fig:llmeas}) lag reset with $p=0.5$ and varying $\gamma$, (\subref{fig:llmeasg}) lag reset with $\gamma=0.5$ and varying $p$.}
\label{fig:OL}
\end{figure}

From the experimental open-loop responses (see {Fig. \ref{fig:OL}}), it appears that the phase margin is larger than predicted for CRONE-1 lag reset with $\gamma=0.5$ and varying $p$, as calculated $\nu$ is expected to give same phase margin.  For CRONE-1 lag reset with $p=0.5$ and varying $\gamma$ the phase is approximately as predicted. Thus some relief from the Bode's gain-phase relation has been achieved. 

In {Fig. \ref{fig:senss}} it was observed that better noise attenuation has been achieved. Contrarily, the open-loop response shows marginal improvement in noise attenuation at frequencies above bandwidth. Also more phase lead is observed around bandwidth. We hypothesize that division between complementary sensitivity and sensitivity function does not relate directly to the open-loop response due to the non-linearity of control. However, this requires further investigation.

\subsection{Reference-tracking}

{\begin{figure}[!htb]
\centering
\includegraphics[width=\linewidth]{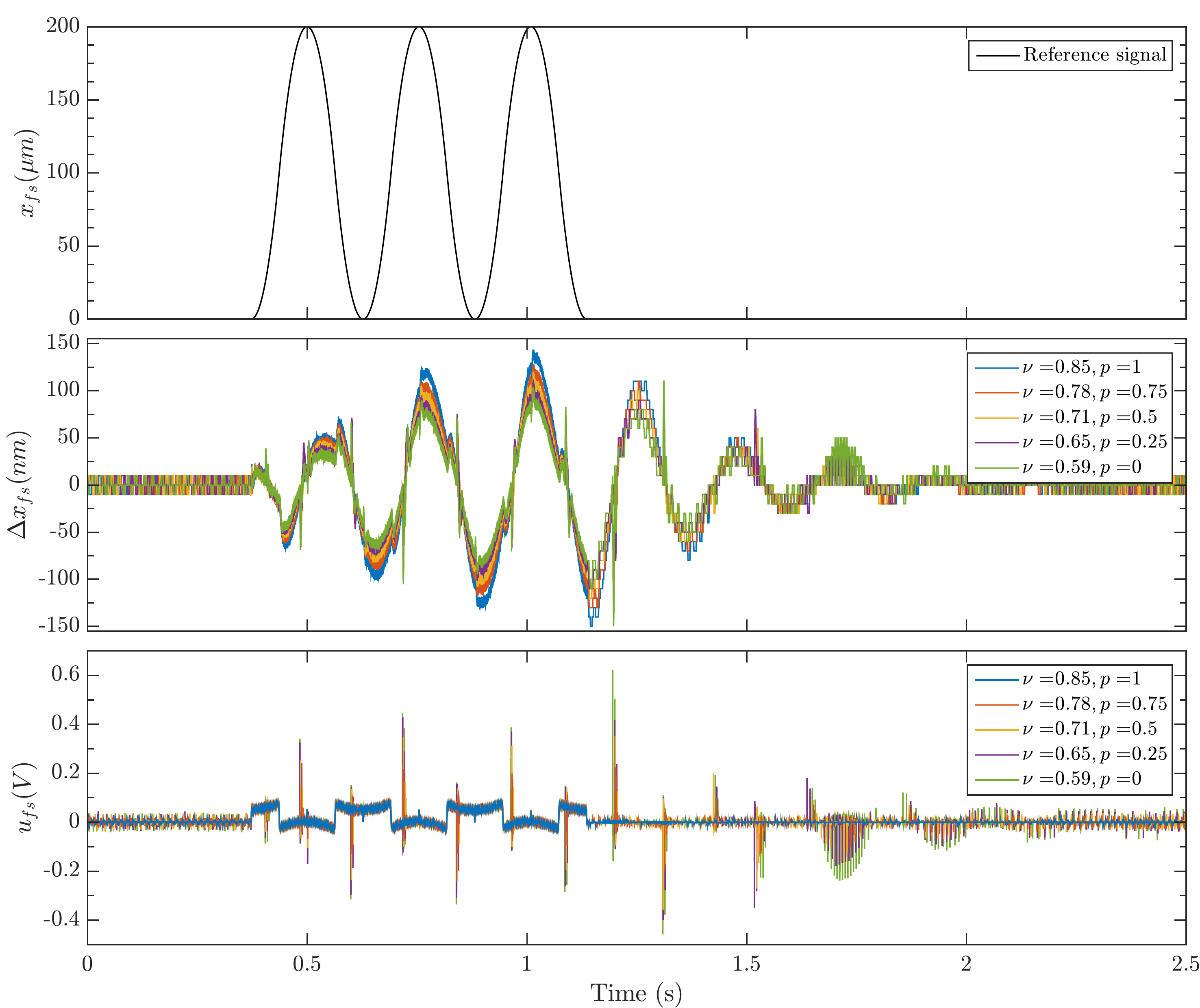}
\caption{Tracking of a triangular wave signal showing the reference signal, tracking error and control input for CRONE-1 lag reset with $\gamma=0.5$ and different values for $p$.}
\label{fig:trackll}
\end{figure}

\begin{table}[!htb]
\centering
\caption{RMS error for fourth order triangular scanning reference CRONE-1 lag reset with constant $\gamma=0.5$}
\label{tab:trackll}
\begin{tabular}{ll}\toprule
$p$&RMS error (\si{\nano\metre})\\\midrule
1&480.1\\
0.75&424.5\\
0.5&385.6\\
0.25&353.4\\
0&331.5\\\bottomrule
\end{tabular}
\end{table}

Fourth order trajectory planning as formulated in	\cite{lambrechts2005} is used to create a triangular wave reference signal. This type of reference signal is representative for scanning motions in precision wafer stages. Also second order feedforward mentioned in the same paper is implemented. The maximum allowed velocity, acceleration and jerk of this reference signal are limited. As can be seen in {Fig. \ref{fig:trackll}}, the tracking error decreases with increasing fractional order $\nu$ as predicted. The RMS error values are listed in Table \ref{tab:trackll} and similarly show better tracking properties when more non-linearity is present in the system. Note that the maximum control input peaks increase  when the system includes reset.
%
\subsection{Noise attenuation}
Time domain response to a sine noise signal of amplitude 2\si{\micro\metre} and frequency of  1\si{\kilo\hertz} is recorded for 5\si{\second} and shown in Fig. \ref{fig:noise1k}. It can be seen that measured output reduces for lower values of $p$ and $\nu$. For a range of frequencies between 300\si{\hertz} and 1\si{\kilo\hertz} similar reduction was observed. The decrease in average power in decibels for $p=0$ and $\gamma=0.5$ with respect to linear case ($p=1$ and $\gamma=0.5$) is shown in Table \ref{tab:noise}. 

\begin{table}[!htb]
\caption{Reduction of average power of noise response for $\gamma=0.5$, $p=0$ with respect to linear case}
\centering
\begin{tabular}{ll}\toprule
Frequency (\si{\hertz})&noise reduction (\si{\decibel})\\\midrule
300&2.85\\
400&2.50\\
500&1.90\\
600&2.03\\
700&0.86\\
800&2.20\\
900&1.84\\
1000&3.44\\\bottomrule
\end{tabular}
\label{tab:noise} 
\end{table}

\begin{figure}[!htb]
\centering
\includegraphics[width=\linewidth]{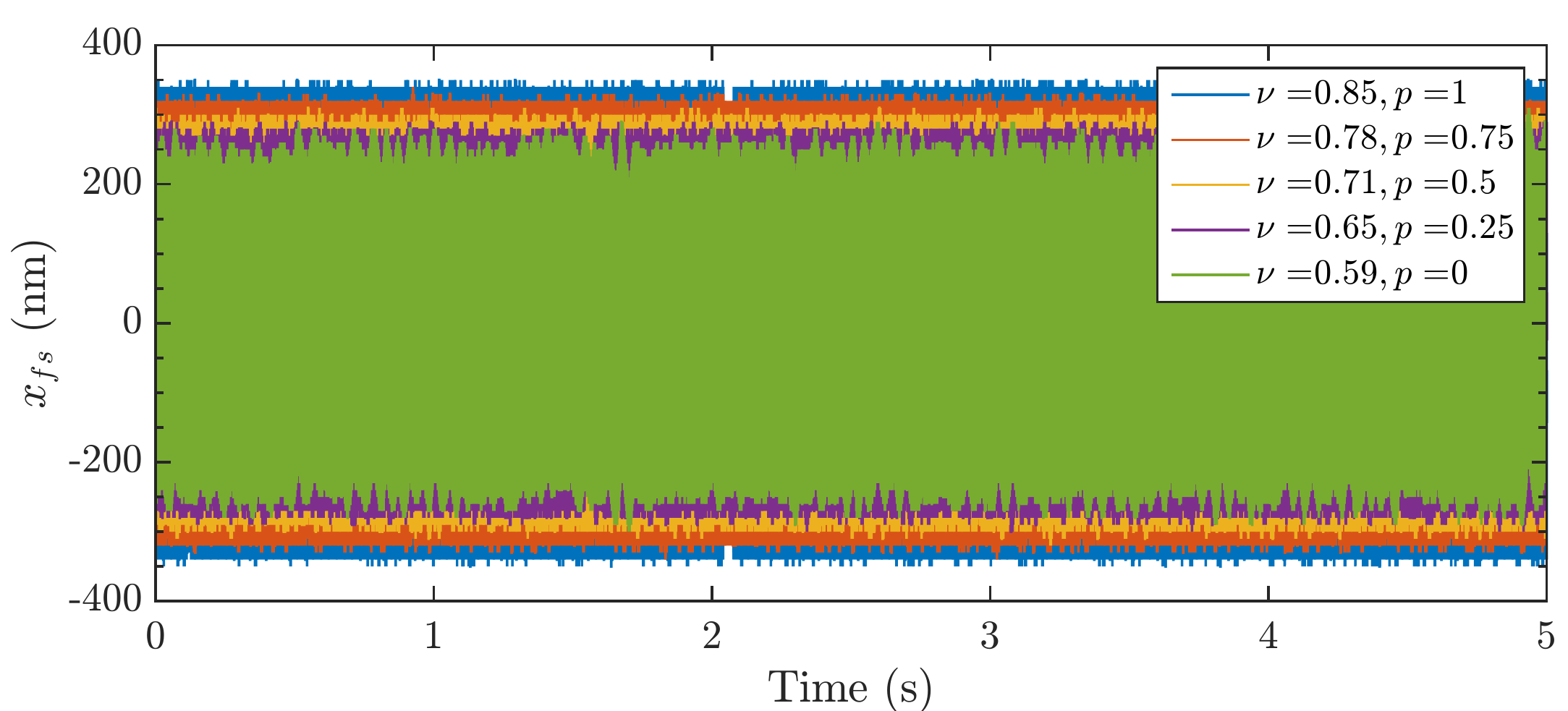}
\caption{Measured response for sine noise signal with frequency of 1\si{\kilo\hertz} and amplitude of 2\si{\micro\metre}.}
\label{fig:noise1k}
\end{figure}

\FloatBarrier
\section{Conclusion}\label{sec:conclusion}
In the first part of this paper, a new design for fractional CRONE control with non-linear reset control was proposed. In the CRONE reset design two degrees of freedom in tuning non-linearity of the system are used to limit the control input peaks and contribution of higher harmonics. 

In the second part of the paper it has been shown that when the additional phase caused by reset can be quantified, and as a result the fractional order $\nu$ of the lag filter can be decreased to the correct value, better system performance is achieved with reset compared to the linear controller for same phase margins. In the open-loop responses that were computed from practical measurements, the responses for CRONE-1 lag reset showed similar phase margins for $\gamma=0.5$ and different $p$-values as well as $p=0.5$ and different $\gamma$-values. As such they show relief from the fundamental Bode's gain-phase relation. However, further investigation of open-loop gain-behaviour in presence of non-linearities is required to explain the absence of significant increase in noise rejection and extra reset phase lead around bandwidth in the obtained open-loop response.  

Several CRONE-1 lag reset controllers, satisfying quadratic stability conditions, have been implemented on a Lorentz-actuated precision stage. On this setup is was shown that tracking error decreased for more non-linearity, implying better reference-tracking of the CRONE reset controller compared to the linear CRONE controller. The identified noise transfer has reduced for frequencies measured in the range from 50\si{\hertz} to 1\si{\kilo\hertz}. This is supported by time domain noise measurements in which transfer of sine signals has been analysed for isolated frequencies. For all chosen frequencies between 300\si{\hertz} and 1\si{\kilo\hertz} noise reduction has been achieved when using CRONE reset instead of CRONE. The retrieved results show that with CRONE reset control relief from fundamental limits in linear control is possible. Having introduced loop shaping methods in the controller design, CRONE reset control is an interesting control method for making reset control more accessible to the industry.

\printbibliography

\end{document}